\documentclass[prd, preprint,tightenlines,floatfix,showpacs,preprintnumbers,
nofootinbib,eqsecnum]{revtex4}
 \usepackage[dvips,final]{graphicx}
  \usepackage{amssymb,amsmath,epsfig,bm,pifont}
   \usepackage{pstricks,pst-node,pst-text,pst-3d}
\usepackage{color}
\def\1{\hbox{{1}\kern-.25em\hbox{l}}}

\begin{document}
\thispagestyle{empty}
 \date{\today}
%\vspace*{-10mm}

\title{Fourth-order QCD 
renormalization group quantities in the {\rm{V}}-scheme   
and the relation of the $\beta$-function to the
Gell-Mann--Low function in QED}
\vspace*{0.8cm}
 \author{A.~L. Kataev}
  \email{kataev@ms2.inr.ac.ru}
   \affiliation{Institute for Nuclear Research of the Academy
of Sciences of Russia, 117312, Moscow, Russia\\}

\author{V. S. Molokoedov }
\email{viktor_molokoedov@mail.ru}
\affiliation{
Moscow Institute of Physics and Technology, 141700 Dolgoprudnyi, Russia\\}
\begin{abstract}
The semi-analytical $O(\alpha_s^4)$ expression for the renormalization 
group $\beta$ function in the ${\rm{V}}$ scheme is obtained in the case 
of the $SU(N_c)$ gauge group. In the process of calculations we use 
the existing information about the   three-loop perturbative  approximation  
for the QCD  
static potential, evaluated in the $\rm{\overline{MS}}$ scheme.
The comparison of the numerical values  
of the third and fourth coefficients  for the QCD  RG $\beta$ functions
in the gauge-independent  ${\rm{V}}$  and  $\rm{\overline{MS}}$ schemes and 
in minimal momentum scheme in the    the Landau gauge   is presented.
The phenomenologically oriented  comparisons  for the coefficients of  
$O(\alpha_s^4)$ expression  
for the   $e^+e^-$-annihilation R-ratio in these schemes are presented.  
It is shown that taking into account these QCD contributions 
is of vital importance and lead to a drastic decrease of the 
scheme-dependence ambiguities of the fourth-order  perturbative 
QCD approximations  for the $e^+e^-$ annihilation R-ratio for the number of 
active flavours, $n_f=5$ in particular.  
We demonstrate that 
in the case of  QED  with $N$-types of leptons  
the coefficients of the   $\beta^{\rm{V}}$ function 
are   closely related to  the ones of the Gell-Mann--Low 
$\Psi$ function and emphasise that they start to differ from each 
other at  the fourth order due to 
the appearance of the  extra $N^2$-contribution
in the V scheme.   
The source  of this  
 extra correction is clarified. 
The general all-order QED relations  
between the   coefficients of the   $\beta^{\rm{V}}$ and $\Psi$ functions      
are discussed.
\end{abstract}
%\Keywords{ QCD, QED, renormalization group; Perturbation theory; V-scheme}
\maketitle

\section{Introduction}

The renormalization group  (RG)   $\beta$ function is one  of the basic  quantities  of  the   
RG method, which was 
developed in the classical works of Refs.\cite{Stueckelberg:1953dz},
\cite{Bogolyubov:1956gh},
\cite{GellMann:1954fq}.
It   defines the energy behaviour of the 
renormalized coupling constants of the renormalized quantum field models. It is 
known  that in the case when the quantum field model under study  
has the single coupling constant,  
the perturbation theory (PT) expressions for its  RG 
$\beta$-functions  depend  on   the choice of the scheme  
of  subtracting ultraviolet (UV) divergences.   

In QED the first and the second  coefficients of the
$\beta$ function are  scheme independent and were   
obtained in   Ref. \cite{GellMann:1954fq} from the analytical calculations  
of the two-loop approximation for the  renormalized photon propagator performed  in  Ref. \cite{JL}.

In the  momentum  (MOM)   scheme,  
defined by  subtractions of   
the UV  divergences  
of the 
photon vacuum polarization function at the non-zero
Euclidean point $\lambda^2$,  the QED   RG $\beta$ function 
coincides with the Gell-Man--Low function
$\Psi(\alpha_{\rm{MOM}})$, where  the  expression for    $\alpha_{\rm{MOM}}(q^2)$ coincides with   the QED   
invariant charge,   uniquely   defined by the combinations of
the  Green functions \cite{Bogolyubov:1980nc}.
The expressions for the coefficients of the  PT series for 
 $\Psi(\alpha_{\rm{MOM}})$ depend on the 
number of leptons $N$.

For   $N$=1, i.e. in the case
of consideration of   the electron only, 
the   three-loop term of the  $\Psi$ function was calculated   analytically 
in Ref. \cite{Baker:1969an}. This  result  was
generalized to the case  of the   arbitrary  number  $N$ of  massless leptons  
in Ref. \cite{Gorishnii:1987fy}. The   $N$-dependent expressions  
for the  four-  and five-loop corrections to the Gell-Man--Low function
were evaluated symbolically    in Refs. \cite{Gorishnii:1990kd} and \cite{Baikov:2012zm} respectively.  At  $N$=1 the result of
Ref. \cite{Baikov:2012zm} coincides with the similar   expression,
obtained in Ref. \cite{Kataev:2012rf}. This feature
should be  considered as the strong  argument in  favour of the  consistency of
the complicated  analytical five-loop  calculations,  
performed in Ref. \cite{Baikov:2012zm}.

Another important scheme, which is used in  QED, is the  on-shell (OS) scheme. 
In this scheme    
the photon vacuum polarization function is 
defined   by subtracting UV  divergences 
at  zero  transferred momentum, while the  renormalized on-shell  masses of
leptons are identified with their experimentally  measured  values.

In the physical OS  scheme the   calculations  of  the
$\beta(\alpha_{\rm{OS}})$ were performed  at the three-loop level in the 
work of   \cite{DeRafael:1974iv}. 
The analytical expression for the corresponding four-loop correction  
was obtained in Ref. \cite{Broadhurst:1992za}. In the case of 
arbitrary $N$ the five-loop contribution   was  obtained   in 
Ref. \cite{Baikov:2012rr}.
It is in agreement with the result of the work \cite{Kataev:2012rf}, 
where this  term was obtained 
at $N$=$1$ with the help of   
the concrete RG-relations.
The agreement with the outcome  of the  direct five-loop  
calculations of  Ref. \cite{Baikov:2012rr} gives extra 
confidence in the correctness and self-consistency of the results of the complicated computer calculations used in 
Ref. \cite{Kataev:2012rf}. 

The third class of  schemes,  which  we are interested in,        
is introduced when 
the dimensional
regularization \cite{'tHooft:1972fi} is used. These schemes 
 include   the  minimal subtractions (MS) 
scheme \cite{'tHooft:1973mm} and its
modified variants, namely the $\rm{\overline{MS}}$-scheme \cite{Bardeen:1978yd}
and the G scheme \cite{Chetyrkin:1980pr}. 
It is possible to prove that for  all   these modifications of
the MS scheme the  coefficients for the  RG  $\beta$ functions
coincide in all orders of PT.  

At $N$=1 the three-loop correction to the  
$\beta(\alpha_{\rm{\overline{MS}}})$ function  was
evaluated in  Refs. \cite{Chetyrkin:1980pr} and \cite{Vladimirov:1979zm} independently  (this result had  
been also  presented  in the review of Ref. \cite{Vladimirov:1979ib}).    
In the case of  the arbitrary  $N$
the three-loop contribution  to  $\beta(\alpha_{\overline{\rm{MS}}})$ 
was  obtained analytically  in Ref. \cite{Gorishnii:1987fy}. The    
computation of the four-loop term was completed  in 
Ref. \cite{Gorishnii:1990kd}. 
The five-loop correction to the QED $\beta$ function in the MS-like schemes  
was  calculated in Ref. \cite{Baikov:2012zm}.
At $N$=1 this expression coincides with the result of non-direct analysis, 
performed in Ref. \cite{Kataev:2012rf}.

It is known that  in QCD  the  MS-like schemes maintain  the explicit 
gauge independence of  various   RG quantities. This property clarifies
why in multiloop QCD  calculations the  MS-like  schemes are used 
more often. In  QCD the  first coefficient of the  $\beta$-function 
was computed in Refs. 
\cite{Gross:1973id},\cite{Politzer:1973fx}
and for the number of quarks flavours $n_f\leq 6$  turned out to be negative.
This feature revealed the existence of the  asymptotic-freedom property
in the gauge   theory of strong interactions.
The two-loop  correction  to the QCD $\beta$ function in the MS-like scheme  
was 
analytically evaluated    in  \cite{Jones:1974mm},
\cite{Caswell:1974gg}, \cite{Egorian:1978zx} and is   also negative \footnote
{Its  first calculation   \cite{Belavin:1974gu} contained a
bug, which resulted
in the  positive value of the two-loop term and in  the appearance  
of the  IR-fixed point of  the  two-loop PT  approximation of  the 
QCD $\beta$ function. This unexpected conclusion stimulated
recalculations of this scheme-independent correction 
\cite{Jones:1974mm}-\cite{Egorian:1978zx} . They resulted  
in  the disappearance of the perturbartive scheme-independent IR-fixed point 
in QCD.}.

At the  three-loop level the  QCD $\beta$ function
was analytically calculated in the $\rm{\overline{MS}}$ scheme in 
Ref. \cite{Tarasov:1980au}. 
This  result 
was  confirmed later in Ref. \cite{Larin:1993tp}. The four-loop
term of the QCD $\beta$ function in the  MS-like schemes  was evaluated  in
Ref. \cite{vanRitbergen:1997va}
and confirmed    in Ref. \cite{Czakon:2004bu}. For  $n_f$=$6$  
the three-loop correction to the $\beta$ function 
in the MS-like schemes  is positive  (see the numerical  results 
presented    below). Note however that  the resummation of the  PT series 
for the QCD $\beta$ function in the MS-like schemes gives 
the arguments that  this feature 
does  not affect the asymptotic freedom  property
\cite{Kazakov:2003df}.  

In QCD one can also use another 
gauge-independent scheme, namely the  V scheme.
It was first  introduced in 
Refs. \cite{Peter:1996ig},
\cite{Schroder:1998vy} and 
is determined by  perturbative   high order QCD corrections 
to the static potential. This scheme  was used in Ref. \cite{Brodsky:1999fr}
to model  massive dependence of the first two coefficients of the  
RG $\beta$ function in the V scheme   and for  the related  
analysis of the manifestation of the  
massive-dependent corrections in the  effect of running of the QCD 
coupling constant from  the  energies above the production
of charm quarks to  
the high energy region above  the scale   $s$=$M_Z^2$. In this case  
the     advantage 
of using the V scheme and not the ${\rm{\overline{MS}}}$ scheme  
is contained 
in the possibility of modelling the  smooth transition of the QCD coupling 
constant through the thresholds of heavy quarks productions. 
Among other applications of the V scheme in QCD  
is the analysis of the perturbative QCD  
predictions for    $\Gamma(H^0\rightarrow b\overline{b})$ 
\cite{Broadhurst:2000yc}. It was shown in this work  that  
within the large $\beta_0$-expansion     
the perturbative approximations   for $\Gamma(H^0\rightarrow b\overline{b})$ 
in the  ${\rm{V}}$ scheme are converging to the concrete stable value 
faster   than in the  ${\rm{\overline{MS}}}$ scheme.

However, to analyse more carefully  the behaviour of various 
perturbative QCD series for the observable physical quantities 
in the V scheme it is necessary   to know high-order PT corrections    
to  the QCD $\beta$ function in this scheme. 
In the present work we will get the semi-analytical result for the 
fourth coefficient  
of the QCD   $\beta$ function in the 
${\rm{V}}$ scheme, i.e. for the 
$\beta^{\rm{V}}$ function.  
In Sec. II the available   
results of the  analytical and semi-analytical 
calculations of the  PT  QCD  corrections to the static 
potential $\rm{V_{QCD}}(\alpha_s(\mu^2))$ in the 
${\rm{\overline{MS}}}$ scheme are summarized. The concrete three-loop 
results, obtained by two groups of authors, are compared.  
Section III is devoted to the definition of the ${\rm{V}}$ scheme
and to the presentation   of the concrete results for the third and 
fourth coefficients of the $\beta^{\rm{V}}$ function. The problem of   
finding the analytical expression for  the concrete known   numerical 
contributions 
to the fourth-order term of  the $\beta^{\rm{V}}$ function is raised. 
In Sec. IV the numerical values for the scheme-dependent coefficients 
of the QCD $\beta$ function in the V scheme 
are compared with  the similar terms, obtained  
in  the MS-like schemes  and in  the  gauge-dependent minimal-MOM (mMOM) scheme widely used at 
present, defined   
in Ref. \cite{vonSmekal:2009ae}. 
We also get the 
$O(\alpha_s^4)$  expression for the $e^+e^-$-annihilation R-ratio 
in the V scheme and compare it with gauge-independent  
 scheme and gauge-dependent  mMOM scheme results.
In fact both $\rm{\overline{MS}}$ and 
 mMOM scheme were applied  
recently for the analysis  of the behaviour of R-ratio 
in the fourth order of PT Ref. \cite{Gracey:2014pba}. 
Using the concrete physical input, we modify this 
analysis and emphasize that it is more consistent
to perform this comparison for $n_f=4,5$ numbers 
of active flavours  in the energy-region 
above the production of charm-quark pairs and below  
$s \approx 900~\rm{GeV^2}$, where the 
effects of subprocess $e^+e^-\rightarrow Z^0\rightarrow {\rm{hadrons}}$ 
did not yet start to manifest itself.   
In Sec. V  we consider the QED limit of the results obtained in Sec. III 
and obtain the expression for the $O(\alpha_{\rm{V}}^5)$ approximation 
of the $\beta^{\rm{V}}$ function in QED. The origin of  difference with 
the QED   Gell-Man--Low  $\Psi$ function,  
which is starting to manifest itself from the 
fourth term, is demonstrated and explained. The  
existing common features of the PT series for the  
$\Psi$ and $\beta^{\rm{V}}$ functions is clarified in all orders of PT.

\section{Preliminaries: the high-order expression of the
static potential in QCD in the $\rm{\overline{MS}}$ scheme}

Let us first   
summarise the available information  about 
the   
perturbative QCD contributions 
to the static potential known at present. This physical quantity 
is used in various   phenomenologically oriented  QCD studies, 
e.g. in the process of  theoretical determinations of the  
charm,  bottom   and top quark masses, and for    
the studies of the   properties of different   
mesons, composed  from  the $c$ and $b$ quarks  
(see e.g.  \cite{Pineda:1998ja},\cite{Ayala:2014yxa},\cite{Kiyo:2014uca}
and  references therein).
 
Within PT the   static potential in QCD is defined 
as a {\it renormalized expression} 
for the  potential of interaction at a  distance {\it r}
 between  static heavy quark $Q_h$ 
and anti-quark $\overline{Q}_h$.
It  is expressed through the  
following Fourier representation  
\begin{eqnarray}
\label{Vq}
 \rm{V}_{QCD}(\mu^2 r^2,\alpha_s(\mu^2))&=
&\int\frac{d^3\vec{q}}{(2\pi)^3}e^{i\vec{q}\vec{r}}{\rm{V}}(\vec{q^2},\mu^2,\alpha_s(\mu^2)) \\ \nonumber 
&=& \int\frac{d^3\vec{q}}{(2\pi)^3}e^{i\vec{q}\vec{r}}
\bigg(-4\pi C_F\frac{\alpha_{s,V}(\vec{q^2}/\mu_V^2)}{\vec{q^2}}\bigg)
\end{eqnarray}
where  
$\alpha_{s,V}(\vec{q^2}/\mu_V^2)$ 
is the renormalized  QCD  coupling constant in 
V scheme, $\alpha_s/4\pi=g^2/16\pi^2$,  
$g$ is the  strong coupling constant of the 
QCD Lagrangian,  
$T^a$ is the  generator of the  $SU(N_c)$ group,
normalized as $T^a=\lambda^a/2$ and $C_F$ is the  Casimir
operator,  defined as $(T^aT^a)_{ij}=C_F\delta_{ij}$.
In the V scheme its coupling constant $\alpha_{s,V}(\vec{q^2}/\mu_V^2)$ 
is related to the numerator
of the momentum representation of the static potential in 
the $\overline{\rm{MS}}$ scheme defined  in Eq. (\ref{Vq})     and is expressed as 
\begin{equation}
\label{av}
\alpha_{s,V}(\vec{q^2}/\mu_V^2)=
\alpha_s(\mu^2)P(\alpha_s(\mu^2),L)=\alpha_s(\mu^2)\sum_{n=0}^{\infty}
P^{\rm{\overline{MS}}}_n(L)\bigg(\frac{\alpha_s(\mu^2)}{4\pi}\bigg)^n~~~.
\end{equation}
The rhs of Eq. (\ref{av}) is expressed  through   higher-order  PT QCD  corrections to   the static potential
$P^{\rm{\overline{MS}}}_n(L)$   
in the $\rm{\overline{MS}}$ scheme which are  known at present   up to  $O(\alpha_s^3)$-level and will be presented below. 

The evolution of the $\rm{\overline{MS}}$ scheme coupling constant  
$\alpha_s(\mu^2)$ (which depends on the $\rm{\overline{MS}}$ scheme  renormalization parameter $\mu^2$)
is governed  by the
QCD  $\rm{\overline{MS}}$ scheme $\beta$ function:
\begin{equation}
\label{betaMS}
\mu^2\frac{\partial(\alpha_s/4\pi)}{\partial\mu^2}=
\beta^{\rm{\overline{MS}}}( a_s)=-\sum_{i=0}^{\infty}\beta_i\bigg(\frac{\alpha_s}{4\pi}\bigg)^{i+2}~~,
\end{equation}
where $a_s=\alpha_s/4\pi$ and   its  known four 
$\rm{\overline{MS}}$ scheme coefficients, 
taken from the work of Ref. \cite{vanRitbergen:1997va}, read:
\begin{eqnarray}
\label{b0}
\beta_0
&=&\frac{11}{3}C_A-\frac{4}{3}T_Fn_l
\\ \label{b1}
\beta_1&=&\frac{34}{3}C^2_A-4C_FT_Fn_l-\frac{20}{3}C_AT_Fn_l
\\ \label{b2}
\beta_2&=&\frac{2857}{54}C^3_A+2C^2_FT_Fn_l-\frac{205}{9}C_FC_AT_Fn_l
-\frac{1415}{27}C^2_AT_Fn_l+
\\ \nonumber
&+&\frac{44}{9}C_FT^2_Fn^2_l
+\frac{158}{27}C_AT^2_Fn^2_l \\ 
\label{b3}
\beta_3&=&\left(\frac{150653}{486}-\frac{44}{9}\zeta(3)\right)C^4_A+
\left(-\frac{39143}{81}+\frac{136}{3}\zeta(3)\right)C^3_AT_Fn_l \\ \nonumber
&+&\left(\frac{7073}{243}-\frac{656}{9}\zeta(3)\right)C^2_AC_FT_Fn_l
+\left(-\frac{4204}{27}+\frac{352}{9}\zeta(3)\right)C_AC^2_FT_Fn_l \\ \nonumber
&+&46C^3_FT_Fn_l+\left(\frac{7930}{81}+\frac{224}{9}\zeta(3)\right)C^2_AT^2_Fn^2_l
\\ \nonumber
&+&\left(\frac{1352}{27}-
\frac{704}{9}\zeta(3)\right)C^2_FT^2_Fn^2_l
+\left(\frac{17152}{243}+\frac{448}{9}\zeta(3)\right)C_AC_FT^2_Fn^2_l
\\ \nonumber
&+&\frac{424}{243}C_AT^3_Fn^3_l+\frac{1232}{243}C_FT^3_Fn^3_l+
\left(-\frac{80}{9}+\frac{704}{3}\zeta(3)\right)\frac{d^{abcd}_Ad^{abcd}_A}{N_A}
\\ \nonumber
&+&
\left(\frac{512}{9}-\frac{1664}{3}\zeta(3)\right)\frac{d^{abcd}_Fd^{abcd}_A}
{N_A}n_l+
\left(-\frac{704}{9}+\frac{512}{3}\zeta(3)\right)
\frac{d^{abcd}_Fd^{abcd}_F}{N_A}n^2_l
\end{eqnarray}
The characteristic colour structures of the group $SU(N_c)$ are defined 
as in the detailed work  of Ref. \cite{vanRitbergen:1998pn}. In the notations 
of Ref. \cite{vanRitbergen:1998pn} we have 
 $[T^a,T^b]=if^{abc}T^c$, where  $f^{abc}$ are the  antisymmetric 
(under permutations of  any pair of indices) structure constants, which  
satisfy the well-known relation
$f^{acd}f^{bcd}=C_A\delta^{ab}$, $C_A$ and $ C_F$ are the  Casimir operators, 
$Tr(T^aT^b)=T_F\delta^{ab}$, $N_A$ is the
number of the  generators of the Lie algebra of the  $SU(N_c)$, 
$n_l$ is  the number of 
quarks flavors, $d^{abcd}_F= Tr(T^aT^{(b}T^cT^{d)})/6 \;$ is the   totally  
symmetric tensor.  
The notations $^{(\dots)}$ are defining the procedure of  symmetrisation of 
the generators $T^{b}T^{c}T^{d}$, 
$d^{abcd}_A=Tr(C^aC^{(b}C^cC^{d)})/6 \;$ is the   total symmetric tensor of 
$(C^a)_{bc}=-if^{abc}$, where  $C^a$ are the generators of the adjoint 
representation 
of the  Lie algebra of the  $SU(N_c)$-group. The  
corresponding  colour structures in   
Eqs.(\ref{b0})-(\ref{b3}) have the following form \cite{vanRitbergen:1997va}:
\begin{eqnarray}
\label{a}
&&C_A=N_c~~~~~,~~~~~C_F=\frac{N_c^2-1}{2N_c}~~~~~,~~~~N_A=N_c^2-1  \\ \label{dA}
&&\frac{d_A^{abcd}d_A^{abcd}}{N_A}=\frac{N_c^2(N_c^2+36)}{24}~,~
\frac{d_F^{abcd}d_A^{abcd}}{N_A}=\frac{N_c(N_c^2+6)}{48}~~\\ \label{dFF}
&&\frac{d_F^{abcd}d_F^{abcd}}{N_A}=\frac{N_c^4-6N_c^2+18}{96N_c^2}~~~~. 
\end{eqnarray}
\newpage 
The terms,  proportional 
to the n-th powers  of  $ L= \ln\left(\mu^2/\vec{q^2}\right)$,  $P^{\rm{\overline{MS}}}_n(L)$ in the polynomial $P(\alpha_s(\mu^2))$  of 
Eq.({\ref{av}) are expressed as  
$P^{\rm{\overline{MS}}}_0=1$, $P^{\rm{\overline{MS}}}_1(L)=
a_1^{\rm{\overline{MS}}}+\beta_0  L$,
$P_2^{\rm{\overline{MS}}}(L)= a_2^{\rm{\overline{MS}}}+(2a_1^{\rm{\overline{MS}}}\beta_0+\beta_1) L+\beta^2_0 L ^2$,
$P_3^{\rm{\overline{MS}}}(L)=a_3^{\rm{\overline{MS}}}+(3a_2^{\rm{\overline{MS}}}\beta_0+2a_1^{\rm{\overline{MS}}}\beta_1+\beta^{\rm{\overline{MS}}}_2) L+(3a_1^{\rm{\overline{MS}}}
\beta^2_0+\frac{5}{2}\beta_0\beta_1){ L}^2+\beta^3_0{L}^3$. The  
 powers of $L$ in the expressions  presented above  
arise from  the solutions of the  corresponding  
RG equations in the MS-like schemes at the three-loop level.

The coefficients $a_i^{\rm{\overline{MS}}}$ are  calculated from the concrete 
Feynman diagrams. The first one,   $a_1^{\rm{\overline{MS}}}$,  
was calculated long time ago  in Refs. \cite{Fischler:1977yf},
\cite{Billoire:1979ih}
and has the following form
\begin{equation}
\label{a1}
a_1^{\rm{\overline{MS}}}=\frac{31}{9}C_A-\frac{20}{9}T_F n_l
\end{equation}
where $n_l=n_f-1$. 
The coefficient $a_2^{\rm{\overline{MS}}}$ was obtained  in the  
\cite{Peter:1996ig}. 
The bug in the pure Yang-Mills contribution to  $a_2^{\rm{\overline{MS}}}$, evaluated  in   Refs. \cite{Peter:1996ig}, 
 was detected 
in Ref. \cite{Schroder:1998vy}\footnote{This correction 
was confirmed  later  by the author in Ref. \cite{Peter:1996ig}}. 

The final result of these  analytical calculations of Refs. \cite{Peter:1996ig},
\cite{Schroder:1998vy} is 
\begin{eqnarray}
\nonumber
a_2^{\rm{\overline{MS}}}&=&\bigg(\frac{4343}{162}+4\pi^2
-\frac{\pi^4}{4}+\frac{22}{3}\zeta(3)\bigg)C^2_A -
\left(\frac{1798}{81}+\frac{56}{3}\zeta(3)\right)C_AT_Fn_l   \\ 
\label{a2}
&-&\left(\frac{55}{3}-16\zeta(3)\right)C_FT_Fn_l+
\left(\frac{20}{9}T_Fn_l\right)^2~~~~.
\end{eqnarray}
 The three-loop  
constant 
perturbative 
contribution to the static
potential in the $\rm{\overline{MS}}$ scheme 
can be presented as
\begin{equation}
\label{a3}
a_3^{\rm{\overline{MS}}}=a^{(3)}_3n^3_l+a^{(2)}_3n^2_l+a^{(1)}_3n_l+a^{(0)}_3~~~.
\end{equation}
The $n_l$-dependent terms were computed  in Ref. \cite{Smirnov:2008pn} and 
have the following form:
\begin{eqnarray}
\label{a33}
a^{(3)}_3&=&-\bigg(\frac{20}{9}\bigg)^3T^3_F
\\ \label{a32}
a^{(2)}_3&=&\left(\frac{12541}{243}+\frac{368}{3}\zeta(3)+\frac{64\pi^4}{135}\right)C_AT^2_F
+\left(\frac{14002}{81}-\frac{416}{3}\zeta(3)\right)C_FT^2_F
\\ \label{a31}
a^{(1)}_3&=&-709.717C^2_AT_F
+\left(-\frac{71281}{162}+264\zeta(3)+80\zeta(5)\right)C_AC_FT_F \\ \nonumber
&+&\left(\frac{286}{9}+\frac{296}{3}\zeta(3)-160\zeta(5)\right)C^2_FT_F-
56.83(1)\frac{d^{abcd}_Fd^{abcd}_F}{N_A}
\end{eqnarray}
where the error of numerical calculation of the   $C_A^2T_F$-coefficient in 
Eq. (\ref{a31}) is not indicated  in Ref. \cite{Smirnov:2008pn}. 

It is worth   emphasizing that in the QED limit with  $C_A$=$0$, 
the analytical expressions of the $n_l$-dependent  terms, 
which are proportional to the  powers 
of $T_F$  
in  Eqs. (\ref{a1}),(\ref{a2}) and in  Eqs. (\ref{a33})-(\ref{a31}),   
are in agreement with the $\rm{\overline{MS}}$-scheme  results presented   in 
\cite{Gorishnii:1991hw}
  for the
constant terms of  the  three-loop approximation of  the
 photon vacuum polarization function in QED. 
They were  also confirmed in Ref. \cite{Baikov:2012rr} 
in the process of  computation  
of the    four-loop approximation of this quantity.      
The  agreement with the  QED results 
of Refs. \cite{Gorishnii:1991hw} 
gives us extra confidence in the validity of the outcomes 
of calculations  of Ref. \cite{Smirnov:2008pn}.

The numerical expressions of the   $n_l$-independent contributions  to  
Eq. (\ref{a3})
were  obtained in Ref. \cite{Smirnov:2009fh}  and read
\begin{equation}
\label{a30}
a^{(0)}_3=502.24(1)C^3_A-136.39(12)\frac{d^{abcd}_Fd^{abcd}_A}{N_A}
\end{equation}
These  results  should be compared  
with the results of the independent calculation 
of Ref. \cite{Anzai:2009tm}
\begin{equation}
\label{anzai}
a^{(0)}_3=502.22(12)C^3_A-136.8(14)\frac{d^{abcd}_Fd^{abcd}_A}{N_A}
\end{equation}
which have  greater inaccuracies.  
Recently the more accurate result for the  second term 
in  Eq. (\ref{anzai}) was obtained
with the help of  the computer code used in  
Ref. \cite{Anzai:2009tm}. 
The improved result for Eq. (\ref{anzai}) is:
\begin{equation}
\label{Sumino}
a^{(0)}_3=502.22(12)C^3_A-136.6(2)\frac{d^{abcd}_Fd^{abcd}_A}{N_A}
\end{equation}
The numerical expression 
of the  coefficient before 
the second structure in  Eq. (\ref{Sumino}) is in  agreement 
with the numerical expression of the same  coefficient in  Eq. (\ref{a30}) 
and demonstrates the reliability of the computer codes, created  in the 
process of calculations, which were 
performed in Ref. \cite{Smirnov:2009fh} and Ref. \cite{Anzai:2009tm}
\footnote{We are grateful to Y. Sumino for informing us of this new  
unpublished result of his personal calculations.}. 
 
The three-loop $n_l$-independent 
correction to the static potential also  contains   the RG non-controllable 
additional   term $8\pi^2C_A^3L$       \cite{Kniehl:2002br}.   It  
is associated with the infrared (IR) divergences, which begin  
to  manifest themselves   
in the  the static potential at the three-loop level 
\cite{Appelquist:1977es}, \cite{Brambilla:1999qa}.  
In the effective theory of heavy quarkonium --
nonrelativistic QCD--
these 
IR-divergent $L$-terms are cancelled by the concrete UV-divergent
contributions (see e.g. \cite{Brambilla:1999qa}).
 
 Among the  aims  of this work is the   
determination of      the four-loop 
approximation of the RG $\beta$ function in the V scheme.  
This can be done by application of the    
RG-motivated    
effective charges (ECH) approach, developed in all orders of PT in  
the  works 
of  Refs. \cite{Grunberg:1980ja},\cite{Grunberg:1982fw}
and independently  at the next-to-leading order (NLO) in 
Ref. \cite{Krasnikov:1981rp} (for the concrete NLO   applications  
see e.g. the work
\cite{Kataev:1981gr}) 
The fourth-order  approximation of  the $\beta$ function in the V scheme 
defines the evolution of $\alpha_{s,V}$ in the region of intermediate and 
UV values of energy scales. It does not depend on the manifestation 
of IR physical effects and on  the RG-uncontrollable 
$L$-dependent corrections to the static potential.  
In view of this we will not consider them 
in our further analysis.     

\section{The fourth order approximation 
of the QCD    $\beta$ function in the V scheme} 
\label{Section:3}
\subsection{The  scale-scheme dependence ambiguities.}

Let us start this section from  writing the RG equation for 
the static potential, which is defined 
in Eq. (\ref{Vq}). 
In the massless limit, considered in this work,  it has 
the following form 
\begin{eqnarray}
\nonumber
\bigg( \mu^2\frac{\partial}{\partial\mu^2}+\beta(a_s)\frac{\partial}{\partial a_s}\bigg)
{\rm{V}}( \vec{q^2}, \mu^2, a_s(\mu^2))=0
\end{eqnarray}

In QCD the scheme-dependence feature   of  the  PT series 
 for the   RG  $\beta$ function  
is the more delicate 
issue than the scheme-dependence problem  of the QED   RG  $\beta$ function  discussed in  the Introduction.  Indeed, contrary to the  QED case, in 
this realistic   theory of strong interactions  it is impossible to 
introduce straightforwardly   
the gauge-invariant 
analog  of the MOM scheme (see e.g. \cite{Celmaster:1979km},
\cite{Braaten:1981dv},\cite{Dhar:1981jm})  
and thus to  construct   
the invariant charge in a unique gauge-invariant manner. 
In QCD the number of the   invariant-type charges  of the  MOM schemes is 
proportional to  4, namely to  the  number of   
vertexes of the Lagrangian 
(i.e. of the  gluon-quark-antiquark, gluon-ghost-ghost, three-gluon and 
the  four-gluon    
vertexes). Moreover,  the definitions of these  invariant-type charges  
depend on  different kinematic conditions for  fixing 
the scales of subtractions of UV divergences in  the  
renormalized Green  functions, which enter these different QCD   invariant-type 
charges.  Indeed, fixing the  kinematics conditions 
by a different way it is possible to construct a number of MOM schemes, 
i.e.  the   
symmetric MOM scheme \cite{Celmaster:1979km}, the   
variant of symmetric MOM scheme  with one external zero momentum 
\cite{Braaten:1981dv} and the asymmetric MOM (AMOM)  scheme  
\cite{Dhar:1981jm}. 
Different gauge-dependent MOM schemes were used in  the 
direct calculations of  the 
massless two-loop \cite{vonSmekal:2009ae}, 
\cite{Raczka:1988my}-\cite{Gracey:2011pf}, 
three-loop \cite{vonSmekal:2009ae}, \cite{Chetyrkin:2000dq}-\cite{Gracey:2011pf}
and even  four-loop   \cite{vonSmekal:2009ae}, \cite{Chetyrkin:2000fd}, \cite{Gracey:2013sca} corrections 
to the QCD $\beta$ function. These analytical calculations  
revealed the importance of the careful study of  the dependence 
on gauge parameter\footnote{It is worth  emphasizing 
 that in the Landau gauge the  two-loop expression  of 
the QCD $\beta$ function in the number of  {\rm MOM} schemes coincide with 
the MS scheme results.}.
 The classical example of the validity of this statement is  the 
discovery that in the AMOM the non-proper choice of the gauge in the 
two-loop PT correction to the QCD $\beta$ function 
can destroy the asymptotic freedom property of the  perturbative QCD 
\cite{Raczka:1988my},\cite{Tarasov:1990ps}.

Summarizing  the discussions of the gauge ambiguities in the QCD analogs 
of the invariant  charges of various MOM schemes,  we stress  that   
in these schemes  it is impossible 
to construct gauge-invariant  analog of the Gell-Man--Low function.   
In view of this it is important  to study the expansions 
of  the  $\beta$ function in terms of   physical  coupling constants, 
which enter the effective LO   approximations of the RG-invariant physical 
quantities, e.g.  the  effective coupling constant of the 
${\rm V}$ scheme defined by the  
QCD  static potential \cite{Brodsky:1999fr}. 

In all these studies the ECH method, developed in  Refs. \cite{Grunberg:1980ja}-\cite{Krasnikov:1981rp},
 was used. To  remind the basis 
of this approach   consider first the system  of  Eq. (\ref{Vq}) and  
(\ref{av}), which defines the expansion of the QCD coupling constant 
of  the ${\rm{ V}}$ scheme through the QCD coupling constant in the 
$\rm{\overline{MS}}$ scheme.
 
At the first step, following the NLO definition of the ECH scheme,  
we  define the effective scale of the ${\rm{V}}$ scheme 
as 
\begin{equation}
\label{scale}
\mu_{V}^2=\displaystyle\rm{exp}[a_1^{\rm{\overline{MS}}}/\beta_0] \mu^2_{\rm{\overline{MS}}}
\end{equation} 
where  $a_1^{\rm{\overline{MS}}}=\frac{31}{9}C_A-\frac{20}{9}T_F n_l$
and $\beta_0$ is the first coefficient of the QCD $\beta$ function, 
defined in Eq. (\ref{betaMS}). 
At the next step we fix 
$\vec{q^2}=\mu_{V}^2$ in Eq. (\ref{av}) and get the following 
relation  between the    effective charge of the ${\rm{V}}$ scheme
and  the QCD coupling constant $\alpha_{s,\overline{\rm{MS}}}$:  
\begin{eqnarray}
\label{asV}
\alpha_{s,V}(\mu_{V}^2)&=&\alpha_{s,\overline{\rm{MS}}}(\mu_{V}^2)
P(\alpha_{s,\overline{\rm{MS}}}, L=0) \\ \nonumber   
&=&\alpha_{s,\overline{\rm{MS}}}(\mu_{V}^2)\bigg[1+a_2^{\rm{\overline{MS}}}
\bigg(\frac{\alpha_{s,\overline{\rm{MS}}}(\mu_V^2)}{4\pi}\bigg)^2
+a_3^{\rm{\overline{MS}}}\bigg(\frac{\alpha_{s,\overline{\rm{MS}}}(\mu_V^2)}{4\pi}\bigg)^3
+O(\alpha_{s,\overline{\rm{MS}}}^4)\bigg]~~.
\end{eqnarray} 

Now it is possible to define the   
ECH $\beta$ function of the  static potential, which is  the 
RG $\beta$ function in the ${\rm{V}}$ scheme  
\begin{equation}
\label{betaV}
\mu_{V}^2\frac{\partial(\alpha_{s,V}/4\pi)}{\partial\mu_{V}^2}=
\beta^{\rm{V}}(a_{s,V})=-\sum_{i=0}^{\infty}\beta_{i}^{V}
\bigg(\frac{\alpha_{s,V}}{4\pi}\bigg)^{i+2}~~~
\end{equation}
where   $a_{s,V}=\alpha_{s,V}/4\pi$. 
The standard RG equation 
relates  $\beta^{\rm{V}}$ function  to  the $\beta$ function in the MS-like schemes:
\begin{equation}
\label{betaVd}
\beta^{\rm{V}}(a_{s,V}(a_{s,\overline{\rm{MS}}}(\mu_{\rm{V}}^2))
=\beta^{\rm{\overline{MS}}}(a_{s,\overline{\rm{MS}}}(\mu_{\rm{V}}^2))\frac
{d a_{s,V}(a_{s,\overline{\rm{MS}}}(\mu^2_{\rm{V}}))}{da_{s,\overline{\rm{MS}}}
(\mu_{\rm{V}}^2)}~~~.
\end{equation}
Consider now the relation between   
$\beta$ functions, computed in the gauge-invariant UV 
subtraction schemes: 
\begin{equation}
\label{relation}
\tilde{\beta}(\tilde{a_s}(a_s))=\beta(a_s)\frac{d\tilde{a_s}
(a_s)}{da_s} .
\end{equation}
where we use the similar normalization 
conditions for both   $\beta(a_s)$ and 
$\tilde{\beta}(\tilde{a_s})$ function, namely 
\begin{equation}
\label{tildebeta}
\mu^2\frac{\partial(\tilde{\alpha_s}/4\pi)}{\partial\mu^2}=
\tilde{\beta_s}(\tilde{a_s})=-\sum_{i=0}^{\infty}\tilde{\beta}_i
\bigg(\frac{\tilde{\alpha_s}}{4\pi}\bigg)^{i+2}~~~.
\end{equation}
with $\tilde{a_s}=\tilde{\alpha_s}/4\pi$.
For these normalization conditions  the coupling constant of one gauge-invariant renormalization   scheme   
$\tilde{a_s}(\mu)$ is related to the  coupling constant   
$\alpha_s(\mu)$ of another gauge invariant renormalization scheme by  the following expression:
\begin{equation}
\label{relation2}
\tilde{\alpha_s}(\mu^2)=\alpha_s(\mu^2)\bigg(1+a_1\bigg(\frac{\alpha_s(\mu^2)}{4\pi}\bigg)+
a_2\bigg(\frac{\alpha_s(\mu^2)}{4\pi}\bigg)^2
+a_3\bigg(\frac{\alpha_s(\mu^2)}{4\pi}\bigg)^3+O(\alpha_s^4)\bigg)~~.
\end{equation}
Taking into account Eq. (\ref{relation}),  the definitions for  
$\tilde{\beta}(\tilde{a_s})$ in Eq. (\ref{tildebeta}) and  the relation   
of Eq. (\ref{relation2}), it is possible to get the following 
links  between the coefficients of the  $\beta$ functions in 
two gauge-invariant schemes:  
\begin{eqnarray}
 \label{t0}
\tilde{\beta}_0
&=&\beta_0
\\  \label{t1}
\tilde{\beta}_1&=&\beta_1
\\   \label{t2} 
\tilde{\beta}_2&=&\beta_2-a_1\beta_1+(a_2-a^2_1)\beta_0
\\    \label{t3v}
\tilde{\beta}_3&=&\beta_3-2a_1\beta_2+a^2_1\beta_1+(2a_3-6a_1a_2+4a^3_1)\beta_0
 \end{eqnarray}
These formulas reflect the transformation laws of the  
$\beta$ function from one gauge-invariant renormalization scheme to another
one. 
\subsection{The  V scheme  $\beta$ function in QCD : Its $O(\alpha_{s,{\rm{v}}}^6)$-approximation.}     
\label{betaV:QCD}
 
Consider now the fourth-order approximation     
of the QCD $\beta$ function in the  V scheme.  
It is related to the    QCD $\beta$ function of the  ${\rm{\overline{MS}}}$ scheme  
via the  Eq. (\ref{betaVd}). Its   
gauge-independent coefficients       
can be obtained from Eqs. (\ref{t0})-(\ref{t3v}), 
where
\begin{eqnarray}
\label{beta0}
\beta^V_0&=&\beta^{\rm{\overline{MS}}}_0=\frac{11}{3}C_A-\frac{4}{3}T_Fn_l~~, 
\\ \label{beta1}
\beta^V_1&=&\beta^{\rm{\overline{MS}}}_1 =\frac{34}{3}C_A^2-4C_FT_Fn_l-
\frac{20}{3}C_AT_Fn_l~~, 
\end{eqnarray}
and $\tilde{\beta}_i=\beta^V_i$, $\beta_i=\beta^{\rm{\overline{MS}}}_i$  with $i=2,3$ and $a_j=a_j^{\rm{\overline{MS}}}$  for $j=1,2,3$. 
Using  the   concrete results   for  
$\beta^{\rm{\overline{MS}}}_i$ (with $i=0,1,2,3$)  
and  $a_j=a_j^{\rm{\overline{MS}}}$ (with $j=1,2,3$) from    
Eq. (\ref{t2}) and (\ref{t3v}) of  Sec. II,   we get  the 
third and fourth coefficients 
$\beta_2^{\rm{V}}$ and $\beta_3^{\rm{V}}$ of the QCD $\beta$ function in the 
V scheme: 
\begin{eqnarray}
\label{b2V}
&&\beta^V_2=\bigg(\frac{206}{3}+\frac{44\pi^2}{3}-\frac{11\pi^4}{12}+
\frac{242}{9}\zeta(3)\bigg)C^3_A 
-\bigg(\frac{445}{9}+
\frac{16\pi^2}{3}-\frac{\pi^4}{3}+\frac{704}{9}\zeta(3)\bigg)C^2_AT_Fn_l \\ 
\nonumber 
&&+2C^2_FT_Fn_l -\left(\frac{686}{9}
-\frac{176}{3}\zeta(3)\right)C_AC_FT_Fn_l+
\left(\frac{2}{9}+\frac{224}{9}\zeta(3)\right)
C_AT^2_Fn^2_l \\ \nonumber
&&+\left(\frac{184}{9}-\frac{64}{3}\zeta(3)\right)C_FT^2_Fn^2_l~~~~; \\    
\label{b3V}
&&\beta^V_3=\bigg(-\frac{5914367}{4374}+\frac{22}{3}\cdot 502.24(1)-
\frac{2728\pi^2}{9}+\frac{341\pi^4}{18}-\frac{15136}{27}\zeta(3)\bigg)C^4_A 
\\ \nonumber
&&+\bigg(\frac{4841537}{2187}-\frac{22}{3}\cdot 709.717
-\frac{8}{3}\cdot502.24(1)+\frac{2752\pi^2}{9}
-\frac{172\pi^4}{9}
+\frac{18184}{9}\zeta(3)\bigg)C^3_AT_Fn_l \\ \nonumber
&+&\left(-\frac{15290}{9}+\frac{1952}{3}\zeta(3)
+\frac{1760}{3}\zeta(5)\right)C^2_AC_FT_Fn_l 
+\left(\frac{572}{9}+\frac{2288}{3}\zeta(3)
-\frac{3520}{3}\zeta(5)\right)C_AC^2_FT_Fn_l \\ \nonumber  
&+&46C^3_FT_Fn_l
+\bigg(-\frac{740860}
{729}+\frac{8}{3}\cdot 709.717-\frac{640\pi^2}{9}+\frac{3208\pi^4}{405}-
\frac{5696}{9}\zeta(3)\bigg)C^2_AT^2_Fn^2_l \\ \nonumber
&+&\left(-\frac{232}{9}-\frac{1024}{3}\zeta(3)
+\frac{1280}{3}\zeta(5)\right)C^2_FT^2_Fn^2_l+
\left(\frac{9328}{9}-448\zeta(3)-\frac{640}{3}\zeta(5)\right)
C_AC_FT^2_Fn^2_l \\ \nonumber
&+&\bigg(\frac{9376}{81}-\frac{512\pi^4}{405}
+\frac{128}{27}\zeta(3)\bigg)C_AT^3_Fn^3_l 
+\left(-128+\frac{256}{3}\zeta(3)\right)C_FT^3_Fn^3_l \\ \nonumber   
&+&\left(-\frac{80}{9}+
\frac{704}{3}\zeta(3)\right)\frac{d^{abcd}_Ad^{abcd}_A}{N_A} \\ \nonumber  
&+&\left(\frac{512}{9}-\frac{1664}{3}\zeta(3)\right)
\frac{d^{abcd}_Fd^{abcd}_A}{N_A}n_l+
\left(-\frac{704}{9}+\frac{512}{3}\zeta(3)\right)
\frac{d^{abcd}_Fd^{abcd}_F}{N_A}n^2_l \\ \nonumber
&-&\frac{22}{3}\cdot 56.83(1)C_A\frac{d^{abcd}_Fd^{abcd}_F}{N_A}n_l
-\frac{22}{3}
\cdot 136.39(12)C_A\frac{d^{abcd}_Fd^{abcd}_A}{N_A} \\ \nonumber
&+&\frac{8}{3}\cdot 56.83(1)\frac{d^{abcd}_Fd^{abcd}_F}{N_A}T_Fn^2_l
+\frac{8}{3}\cdot 136.39(12)\frac{d^{abcd}_Fd^{abcd}_A}{N_A}T_Fn_l~~~~.
\end{eqnarray}

The property of the  {\it scheme independence} of
the coefficients  $\beta^V_{i}$ within the gauge-independent MS-like   schemes 
is the consequence of application of the ECH  
approach to the static potential. Indeed, it is possible 
to show that  these coefficients  
are  related  to  the massless gauge-independent scheme invariants, 
 introduced in the work of Ref. \cite{Stevenson:1981vj} (for the  details 
of derivation see e.g. Ref. \cite{Kataev:1995vh}).  
The analytical  expression  for Eq. (\ref{b2V}) was 
obtained in 
Ref. \cite{Schroder:1998vy} and agrees with the similar  
one of Ref. \cite{Peter:1996ig} with the $C_A^3$-term corrected later on. 

The  result   of Eq. (\ref{b3V}) is 
new.   Its semi-analytical form is explained by 
the similar representation presented in Sec. II  for the  
coefficients of Eq. (\ref{a31}), (\ref{a30}) and of  
Eqs. (\ref{anzai}), (\ref{Sumino}), obtained 
in the works \cite{Smirnov:2008pn}, 
\cite{Smirnov:2009fh} and by the authors of Ref. \cite{Anzai:2009tm} 
respectively. 

Consider now the  real QCD case, based on the  $SU(N_c$=$3)$  
gauge group of colour. In the fundamental representation its group 
structures are fixed as  $C_A=3$, $C_F=4/3$, $T_F=1/2$, $N_A=8$, 
$d_A^{abcd}d_A^{abcd}=135$, 
$d_F^{abcd}d_A^{abcd}=15/2$ and $d_F^{abcd}d_F^{abcd}=5/12$. 
Converting  now  the $SU(N_c)$-group   expressions presented above   for the   
coefficients of the QCD    $\beta^{\rm{V}}$ function  
into the form corresponding to 
the $SU(3)$ group, we get
the well-known  results for  $\beta_0$ and $\beta_1$         
\begin{eqnarray}
\label{b0-b1}
\beta_0&=&11-0.666666n_l~~~,  \\ \label{B1}
\beta_1&=&102-12.66666n_l~~~,  
\end{eqnarray} 
and the following numerical  expressions  
for the third and fourth coefficients 
of the QCD $\beta^{\rm{V}}$ function : 
\begin{eqnarray}
\label{beta2V}
\beta_2^{V}&=&4224.181-746.0062n_l+20.87191n_l^2  ~~~, \\ \label{beta3V}
\beta_3^{V}&=&43175.06(6.43)-12951.700(390)n_l + 706.9658(6)n^2_l 
- 4.87214n^3_l~~~.
\end{eqnarray}
The errors of the first three terms  in 
Eq.(\ref{beta3V})  are defined as the mean square error 
$\sigma=\sqrt{\sum_{i=1}^{k}\sigma_i^2}$, where $\sigma_i$ are  
the numerical errors that arise from the multiplication 
of the factor $2\beta_0$ by the 
computed errors of  the corresponding  $\rm{\overline{MS}}$-scheme 
numbers 
for  $a^{(1)}_3$ and $a^{(0)}_3$, given in  Eqs. (\ref{a31}) and (\ref{a30}).

\subsection{The guess about analytical representation 
of the numerical terms  in   the $SU(N_c)$ expression  
for    $\beta_3^{V}$ }
 
It may be inspiring to make a guess on the possible analytical 
representations of the results of numerical calculations 
of the concrete terms in $a_3^{(1)}$ and $a_3^{(0)}$ coefficients.
There is the general rule that the rate of transcendentality 
structure   is increasing with increasing order of PT calculations.    

Following this general  rule and considering 
the terms  in the expressions for $\beta_2^{\rm{V}}$  and 
$\beta_3^{\rm{V}}$,  we claim that  
the numerically  evaluated   contributions in the expressions for the 
$a_3^{(1)}$ and $a_3^{(0)}$ coefficients, which enter 
the expressions for the concrete terms in $\beta_3^{\rm{V}}$,  can be decomposed 
in terms of rational and transcendental numbers in the following way:  
\begin{eqnarray}
\label{1}
&&709.717=R_1+R_2\pi^2+R_3\pi^4+R_4\zeta(3)+R_5\pi^2\zeta(3)+R_6\zeta(5) \\
\label{2}
&&502.24(1)=R_7+R_8\pi^2+R_9\pi^4+R_{10}\zeta(3)+R_{11}\pi^2\zeta(3)+R_{12}\zeta(5) \\
\label{3}
&&56.83(1)=R_{11}+R_{12}\pi^2+R_{13}\pi^4+ R_{14}\zeta(3) \\ \label{4}
&&136.39(12)= R_{15}+R_{16}\pi^2+R_{17}\pi^4+ R_{18}\zeta(3) 
\end{eqnarray}
where $R_i$ are still  unknown rational numbers. Note  that the  rational
number is any number that can be expressed as the  ratio 
$\rm{(p/q)}$ of two integers  with non-zero $\rm{q}$.
Thus, some of $R_i$ coefficients  in Eqs. (\ref{1})-(\ref{4}) may be zero. 
There are indications that  $R_{12}$ and $R_{16}$  may really be  zero. 
It will be interesting  to check this guess by analytical calculations 
of the corresponding complicated Feynman diagrams.

\section{The applications 
of the ${\rm{V}}$ scheme in perturbative QCD and 
the results  obtained in the    $\rm{\overline{MS}}$ scheme
and the  minimal $\rm{MOM}$ scheme} 
\label{sec:Comparison}
\subsection{General discussions}
\label{subsec:gd}

In the last few years the interest in studying  the 
 perturbative  expressions  
for  the QCD $\beta$ function in the gauge-independent and gauge-dependent 
schemes increased. This  interest was pushed ahead by the considerations   
of the purity   of the conformal windows  related to the 
IR fixed points  in the expressions for 
the $\beta$ functions of the strong interactions theories, based on the 
concrete non-Abelian groups with fermions (see e.g.    
\cite{Ryttov:2013ura}-\cite{Gracey:2015uaa}).

There are also  more  phenomenologically motivated studies 
of the behaviour of 
various   PT QCD contributions    
to the      RG-invariant  quantities, evaluated     in the  
different UV-subtraction  schemes.  
The  first study of the gauge-dependence of the  three-loop 
corrections to the $e^+e^-$-annihilation  
R-ratio was made within the AMOM scheme in 
Ref. \cite{Raczka:1989wp}. However, this work was based on the analysis of the  
gauge-dependence of the AMOM version of the   
$O(\alpha_s^3)$ contribution  
to  this quantity containing the bugs, evaluated     
in the $\rm{\overline{MS}}$ scheme in Ref. \cite{Gorishnii:1988bc}. 
It is worth recalling that this $\rm{\overline{MS}}$-scheme  result was 
corrected in  Ref. \cite{Gorishnii:1990vf} and confirmed in 
Ref. \cite{Surguladze:1990tg} and later on in Ref. \cite{Chetyrkin:1996ez}.
In view of this it may be  interesting  to 
clarify the status of the gauge dependence of the available   
$O(\alpha_s^4)$ approximation for the $e^+e^-$-annihilation    
R-ratio in the AMOM-scheme using 
the $O(\alpha_s^4)$  
corrections, evaluated recently   in Refs.  \cite{Baikov:2008jh},\cite{Baikov:2012zn}

Quite recently a similar analysis was done at the three-loop 
level in different gauge-dependent MOM schemes, and at the four-loop 
order in the mMOM scheme, specified 
for the case of   the Landau gauge  \cite{Gracey:2014pba}.
This mMOM scheme was formulated   
in Ref. \cite{vonSmekal:2009ae} and already used 
in the theoretical studies of the behaviour of the  gauge-dependent   
QCD $\beta$ function for different numbers of 
fermions flavours $n_f$ (see the works of Refs. \cite{Gracey:2013sca},
\cite{Ryttov:2013ura},\cite{Gracey:2015uaa}). In this 
section we will compare the expressions for the  coefficients of the RG $\beta$ function 
in the V scheme obtained  in 
Sec. \ref{Section:3} with the similar mMOM-scheme   results. 
In the next section we will use the results of 
Sec. \ref{Section:3}
 to  
study  the third and fourth order approximations  of  the $e^+e^-$-annihilation 
R-ratio in the ${\rm{V}}$ scheme  and compare 
it  with the results obtained  in the $\rm{\overline{MS}}$ scheme
and in the mMOM  scheme, which were 
presented in  Ref. \cite{Gracey:2014pba}.

\subsection{The definition of the minimal MOM scheme.}

Let us first briefly  review how the  mMOM scheme is defined.  Using 
the standard notations for the renormalization constants of QCD 
in an arbitrary linear covariant gauge  
namely   
\begin{equation}
\label{not}
\psi_0=\sqrt{Z_{\rm{\psi}}} \psi, \; A^{a\mu}_0=\sqrt{Z_{\rm{A}}} A^{a\mu}, \; c^a_0=\sqrt{Z_{\rm{c}}} c^a,\;  g_0=Z_{\rm{g}}g, \; \lambda_0=Z_{\rm{A}} Z^{-1}_{\lambda}\lambda
\end{equation}
where $\psi, \; A^a_\mu, \; c^a$ are the  quarks, gluons and ghosts fields  
respectively, $g$ is the constant of the strong interaction, 
$\lambda$ is the gauge parameter, which is included in the Lagrangian QCD \
as $(\partial_\mu A^a_{\mu})^2/2\lambda$. We first write down the non-renormalized 
gluon propagator in the  momentum space:
\begin{eqnarray}
D^{\mu\nu}_{ab}=\frac{i\delta_{ab}}{p^2+i\varepsilon}\bigg(-g^{\mu\nu}+(1-\lambda)\frac{p^{\mu}p^{\nu}}{p^2+i\varepsilon}\bigg)~~~~.
\end{eqnarray}
The form of the QCD  Lagrangian dictates how to 
relate different   renormalization constants. 
For example, the  renormalization constant of 
the gluon-ghost-ghost vertex has the following form:   
\begin{equation}
\label{Z}
Z_{\rm{ccg}}=Z_{\rm{g}} Z^{1/2}_{\rm{A}} Z_{\rm{c}}
\end{equation}
The definition of the   mMOM scheme is based on the consideration of 
this relation  \cite{vonSmekal:2009ae}. Taking into account 
Eq. (\ref{Z}) one can write down the expression for  
the QCD coupling constant of  the mMOM scheme
$\alpha_s^{\rm{mMOM}}$ as 
\begin{equation}
\label{mMOM}
\alpha_s^{\rm{mMOM}}(\mu^2)=\frac{Z_{\rm{A}}^{\rm{mMOM}}(\mu^2)
(Z_{\rm{c}}^{\rm{mMOM}}(\mu^2))^2}
{(Z_{\rm{ccg}}^{\rm{mMOM}}(\mu^2))^2}\alpha^0_{ s}~~~~~~. 
\end{equation}
Following the proposals  of Ref. \cite{vonSmekal:2009ae}    
the renormalization expressions for 
the gluon and ghost propagators  are defined by 
using   the requirements that  
at $p^2=\mu^2$ their residues  are equal to unity, namely  
\begin{equation} 
\label{definition}
D(p^2,\alpha_s^{\rm{mMOM}}(\mu^2) )|_{p^2=\mu^2}=1~~,~~
G(p^2,\alpha_s^{\rm{mMOM}}(\mu^2) )|_{p^2=\mu^2}=1~~~.
\end{equation}
Then   the renormalized expression 
 for the gluon propagator, 
defined in the Landau gauge $\lambda$=0, will 
take the following form  
\begin{eqnarray}
D^{\mu\nu}_{ab}&=&i\delta_{ab}\bigg(g^{\mu\nu}-\frac{p^{\mu}p^{\nu}}{p^2+i\varepsilon}\bigg)
\frac{D(p^2,\alpha_s^{\rm{mMOM}}(\mu^2))}{p^2+i\varepsilon} 
\end{eqnarray}
while the expression for the ghost propagator is defined as
 \begin{eqnarray}
D^{c}_{ab}&=&i\delta_{ab}\frac{G(p^2,\alpha_s^{\rm{mMOM}}(\mu^2))}{p^2+i\varepsilon}~~~.
\end{eqnarray}
The most important additional requirements  
of the mMOM scheme \cite{vonSmekal:2009ae},\cite{Gracey:2013sca}  are  
the special  
definitions of the renormalization 
constant of the  gluon-ghost-ghost vertex and of the  renormalization 
constant of the gauge parameter,    
 namely 
\begin{equation}
\label{condition}
Z_{\rm{ccg}}^{\rm{mMOM}}(\alpha_s^{\rm{mMOM}})=Z_{\rm{ccg}}^{\rm{\overline{MS}}}                    
(\alpha_s^{\rm{\overline{MS}}})~~,~~~Z_{\lambda}^{\rm{mMOM}}(\alpha_s^{\rm{mMOM}})=
Z_{\lambda}^{\rm{\overline{MS}}}(\alpha_s^{\rm{\overline{MS}}})~~~.
\end{equation}
Taking into account the definition of the QCD coupling constant 
in the $\rm{\overline{MS}}$ scheme through the same vertex 
\begin{equation}
\label{MS}
\alpha_s^{\rm{\overline{MS}}}(\mu^2)=\frac{Z_{\rm{A}}^{\rm{\overline{MS}}}(\mu^2)
(Z_{\rm{c}}^{\rm{\overline{MS}}}(\mu^2))^2}
{(Z_{\rm{ccg}}^{\rm{\overline{MS}}}(\mu^2))^2}\alpha^0_{s}
\end{equation} 
and  Eqs. (\ref{Z}) and (\ref{condition}),
one can get the 
useful 
relations between the renormalization constants of the mMOM and 
$\rm{\overline{MS}}$ schemes 
\begin{equation}
\label{relations}
Z^{\rm{mMOM}}_{\rm{g}}\sqrt{Z^{\rm{mMOM}}_{\rm{A}}}Z^{\rm{mMOM}}_{\rm{c}}=
Z^{\rm{\overline{MS}}}_{\rm{g}}
\sqrt{Z^{\rm{\overline{MS}}}_{\rm{A}}}Z^{\rm{\overline{MS}}}_{\rm{c}}.
\end{equation}
and the relation between the renormalized QCD coupling constants of 
these schemes
\begin{equation}
\label{alpha}
\alpha^{\rm{mMOM}}_s(\mu^2)=\frac{Z^{\rm{mMOM}}_{\rm{A}}}{Z^{\rm{\overline{MS}}}_{\rm{A}}}\biggl(\frac{Z^{\rm{mMOM}}_{\rm{c}}}{Z^{\rm{\overline{MS}}}_{\rm{c}}}
\biggl )^2 \alpha^{\rm{\overline{MS}}}_s(\mu^2) ~~~~~~~.
\end{equation} 
All formulas written above  are valid for any linear covariant gauge and 
for the  Landau gauge $\lambda$=0  in particular. This choice of the gauge 
leads to the simplification of the final perturbative  results 
we will be interested in. Note also that the application of the 
Landau gauge allows us to  simplify  definite lattice Yang-Mills studies 
(see e.g. \cite{Greensite:2006ns}). 

\subsection{Comparison of the fourth-order  approximations of the 
QCD $\beta$ function in the ${\rm{V}}$,  mMOM and  
$\rm{\overline{MS}}$ schemes.} 
\label{betaVs}
The analytical expressions for 
the three- and four-loop  
coefficients of the QCD $\beta$ function  in the mMOM scheme  
in the general covariant gauge were
 obtained in  Ref. \cite{vonSmekal:2009ae}. In the process of their 
derivation  the $\rm{\overline{MS}}$-scheme results of 
Refs. \cite{vanRitbergen:1997va},\cite{Czakon:2004bu}, supplemented with
 the 
explicit expressions for the relation of Eq. (\ref{alpha}),  and with 
the  
three-loop anomalous dimension of the  gauge parameter 
in the $\rm{\overline{MS}}$ scheme , evaluated    
in   Ref. \cite{Chetyrkin:2000dq}, were used.  
The  results of Ref. \cite{vonSmekal:2009ae} were  
confirmed  recently  in Ref. \cite{Gracey:2013sca}  by direct 
symbolical three- and four-loop   computations. 
In the Landau gauge they take   the following numerical form
\begin{eqnarray}
\label{beta2MoM}
\beta_2^{\rm{{mMOM},\lambda=0}}&=&3040.482-625.3867n_l+19.38330n_l^2 \\ \label{beta3mMoM}
\beta_3^{\rm{{mMOM},\lambda=0}}&=&100541.05-24423.330n_l+1625.4022n_l^2-27.49263n_l^3
\end{eqnarray}
It is interesting to compare these results with the 
numerical expressions of the  same  
coefficients of  the QCD $\beta$ function in the gauge-invariant  
V scheme 
(see Eqs. (\ref{beta2V}) and (\ref{beta3V})) and in the  gauge-invariant 
$\rm{\overline{MS}}$ scheme, namely with  
\begin{eqnarray}
\label{b2n}
\beta_2^{\rm{\overline{MS}}}&=&1428.500-279.6111n_l+6.01851n^2_l~~~, \\ \label{b3n}
\beta_3^{\rm{\overline{MS}}}&=&29242.96-6946.289n_l+405.0890n^2_l+1.49931n^3_l~~~, 
\end{eqnarray}
which follow from the results 
of analytical calculations  of Refs. \cite{Tarasov:1980au} and   
\cite{vanRitbergen:1997va}.

For the  completeness,  in Table 1  we present this comparison 
for all numbers of quarks flavours $1\leq n_f\leq 6$, where $n_f=n_l+1$.
These notations are identical to the ones used 
for fixing the numbers of heavy flavours, which are considered  in the PT  
QCD expression for the static potential V,   
where $n_l$ is 
 the number of quarks, lighter than  $Q_h$. They   enter virtual 
corrections among the  heavy quark 
and antiguark of the  flavour $n_f$ and   vary    
in the  region $3\leq n_l\leq 5$.

\begin{center}
\begin{tabular}{|c|c|c|c|c|c|c|}
\hline
\multicolumn{7}{|c|}{\textbf{ The  numerical coefficients of the QCD 
$\beta$ function in different schemes}} \\
\hline 
$n_f$ & $\beta^{\rm V}_2$  & $\beta^{\rm V}_3$ & $\beta^{\rm{\overline{MS}}}_2$ & 
$\beta^{\rm{\overline{MS}}}_3$ & $\beta^{\rm{mMOM},\lambda=0}_2$ & $\beta^{\rm{{mMOM},\lambda=0}}_3$ \\
\hline 
1 & $\;$ 3499.047 $\;$ &  $30925.46\pm 6.44$  & $\;$ 1154.907 $\;$ & $\;$ 22703.26 $\;$ & $\;$ 2434.478 $\;$
& $\;$ 77715.63 $\;$ \\
\hline
2 & 2815.656 & $20060.55\pm 6.48$ & 893.351 & 16982.73 & 1867.242 
& 57976.06 \\
\hline
3 & 2174.010 & $10551.11\pm 6.54$ & 643.833 & 12090.37 & 1338.771 
& 41157.38 \\
\hline
4 & 1574.107 & $2367.90\pm 6.62$ & 406.351 & 8035.18 & 849.068 & 
27094.64 \\
\hline
5 & 1015.948 & $-4518.30\pm 6.72$ & 180.907 & 4826.15 & 398.131 & 
15622.88  \\
\hline
6 & 499.533 & $-10136.74\pm 6.84$ & -32.500 & 2472.28 & -14.038 & 
6577.14 \\
\hline
\end{tabular}
\vspace{0.2cm}\\
{Table 1. The comparison of the numerical values of  the 
third and fourth coefficients of the QCD $\beta$ function in the 
V , $\rm{\overline{MS}}$ and mMOM scheme in the Landau gauge.}
\end{center}

The results of this Table  demonstrate  that the asymptotic 
structure of the PT series for the  effective $\beta$ function in the 
$\rm{V}$ scheme has the  non-regular behaviour and differs from 
the asymptotic structure of the  PT for the $\beta$ function in the 
$\rm{\overline{MS}}$ scheme, which was  considered  in 
Ref. \cite{Suslov:2005zi}  
using the approach developed in Ref. \cite{Lipatov:1976ny}. 
In view of this 
it is of interest whether  this non-regular behaviour of the PT series 
for the $\beta^{\rm{V}}$ function   will 
manifest itself in the process of studies of scheme dependence 
of high-order coefficients  for the characteristics of 
typical  physical QCD  processes, e.g.  
for the $e^+e^-$-annihilation R-ratio in the region of 
direct production of the pair of  heavy quarks and antiquarks
with $n_f=4,5$ numbers of flavours.
We will  not consider in this work  
the case of $n_f=6$, related to  the direct production 
of the pair of $t\overline{t}$-quarks in the process
$e^+e^-\rightarrow {\rm{hadrons}}$, which may be studies in future 
if the ILC will be built. Indeed, the total cross section  of this process 
is dominated by the subprocess 
 $e^+e^-\rightarrow Z^0\rightarrow  {\rm{hadrons}}$ and not by 
the subprocess  $e^+e^-\rightarrow \gamma \rightarrow  {\rm{hadrons}}$
that interests us  in this work.

\subsection{The fourth-order approximation  for 
the $e^+e^-$ 
R-ratio in the $\rm{\overline{MS}}$ and V schemes}     
\label{SubV}

We  now discuss  the fourth-order PT expression for 
the $e^+e^-$-annihilation R-ratio in the V scheme. 
The idea  to  study  this particular expression, as well 
as the PT expressions  for other observable physical quantities 
in the V scheme, was proposed   some time ago in 
Ref. \cite{Brodsky:1994eh}. In this section we will  realise  
this proposal,  obtain  the fourth-order V scheme    
PT approximation for the $e^+e^-$-annihilation 
ratio $R(s)$ and compare its coefficients and  energy dependence 
with the results, obtained  in the 
$\rm{\overline{MS}}$ scheme and in the Landau gauge variant of the  
mMOM scheme \cite{Gracey:2014pba}.
The studies 
to be made in this subsection  supplement the ones 
presented above. Moreover, the results obtained in  Sect. IV C  
will be used in the process of the  numerical calculations to be 
presented below.  

We remind the reader that the $e^+e^-$-annihilation   R-ratio is 
defined as  
\begin{equation}
\label{def}
R(s)=\frac{\sigma(\rm{e}^{+}\rm{e}^{-} \rightarrow \gamma \rightarrow \text{hadrons})}
{\sigma_0(\rm{e}^{+}\rm{e}^{-}\rightarrow\gamma \rightarrow    
\rm{\mu}^{+}\rm{\mu}^{-})}=12\pi\rm{Im} \; \Pi (s+i\varepsilon)
\end{equation}
where  $s$ is the transferred  energy in the Minksowskian region,
$\sigma_0(\rm{e}^{+}\rm{e}^{-}\rightarrow \gamma \rightarrow 
 \mu^{+}\mu^{-})=
4\pi^2\alpha/(3s)$ is the theoretical normalization factor, 
$\Pi(q^2)$ is the QCD expression for the  
photon vacuum polarization function   
\begin{equation}
\Pi_{\mu\nu}(q^2)=(q_\mu q_\nu-g_{\mu\nu}q^2)\Pi(q^2)=
i\int d^4x \; e^{iqx}\langle 0 | T \; j_\mu(x)j_\nu(0)|0\rangle
\end{equation}  
and  $j_\mu=\sum\limits_f Q_f\bar{\psi}_f\gamma_\mu\psi_f$ 
is the electromagnetic hadronic  current. Since   the $e^+e^-$-annihilation  
R-ratio is the RG-invariant  quantity,  it obeys the RG equation 
without anomalous dimension term, namely  
\begin{equation}
\left( \mu^2\frac{\partial}{\partial \mu^2}+\beta(a_s)\frac{\partial}{\partial a_s}\right) R(s)=0 \; ,
\end{equation}
In the $\rm{\overline{MS}}$ scheme 
the $O(\alpha_s^4)$ approximation 
for the $e^+e^-$  R-ratio has the following form   
\begin{equation}
\label{MSbar}
R^{\rm{\overline{MS}}}=3\sum_f Q^2_f\left(1+4\frac{\alpha^{\rm{\overline{MS}}}_s}{4\pi}+r^{\rm{\overline{MS}}}_1\left(\frac{\alpha^{\rm{\overline{MS}}}_s}{4\pi}\right)^2+r^{\rm{\overline{MS}}}_2\left(\frac{\alpha^{\rm{\overline{MS}}}_s}{4\pi}\right)^3+r^{\rm{\overline{MS}}}_3\left(\frac{\alpha^{\rm{\overline{MS}}}_s}{4\pi}\right)^4\right)
\end{equation}
where the coefficient $r^{\rm{\overline{MS}}}_1$ 
was evaluated analytically in Ref. \cite{Chetyrkin:1979bj} and 
numerically in 
Ref. \cite{Dine:1979qh} and confirmed analytically  in 
Ref. \cite{Celmaster:1979xr}. The coefficient 
$r^{\rm{\overline{MS}}}_2$ was analytically evaluated in 
Ref. \cite{Gorishnii:1990vf} and confirmed in   
Refs. \cite{Surguladze:1990tg} and \cite{Chetyrkin:1996ez}, while the symbolical expression for the {\it non-singlet} 
and {\it singlet} contributions to  $r^{\rm{\overline{MS}}}_3$
were obtained analytically only recently   in  
Ref. \cite{Baikov:2008jh} and \cite{Baikov:2012zn}
respectively. 
The coefficients   $r^{\rm{\overline{MS}}}_1$,
$r^{\rm{\overline{MS}}}_2$ and  $r^{\rm{\overline{MS}}}_3$ can be expressed 
in the  
numerical form as  
\begin{eqnarray}
\label{r1}
r^{\rm{\overline{MS}}}_1&=&-1.84472n_f+31.7713 \; , \\ \label{r2}
r^{\rm{\overline{MS}}}_2&=&-0.33139n^2_f-76.8085n_f-424.763-26.4435\delta_f \; , \\ 
\label{r3}
r^{\rm{\overline{MS}}}_3&=&5.50812n^3_f-204.1431n^2_f+4806.339n_f-40091.67 \\ 
\nonumber
&+&(49.0568n_f-1521.214)\delta_f .
\end{eqnarray}
where the terms, proportional to 
$\delta_f=(\sum_f Q_f)^2/(\sum_f Q^2_f)$, are the {\it singlet} 
contributions.  

In the  V scheme the  PT expression for the $e^+e^-$ 
R-ratio is defined as 
\begin{equation}
R^{\rm{V}}=3\sum_f Q^2_f\left(1+4\frac{\alpha_{s,\rm{V}}}{4\pi}+r^{\rm{V}}_1\left(\frac{\alpha_{s, \rm{V}}}{4\pi}\right)^2+r^{\rm{V}}_2\left(\frac{\alpha_{s, \rm{V}}}{4\pi}\right)^3+r^{\rm{V}}_3\left(\frac{\alpha_{s, \rm{V}}}{4\pi}\right)^4\right)
\end{equation}
Using the ECH approach of Ref. \cite{Grunberg:1982fw} 
and the V scheme relations of Eqs. (\ref{scale}),  (\ref{t2}) and 
 (\ref{t3v}) we obtain the following  general expressions 
for $r^V_i$:
\begin{eqnarray}
\label{r1VG}
r^{\rm{V}}_1&=&r^{\rm{\overline{MS}}}_1-4a^{\rm{\overline{MS}}}_1 , \\ \label{r2VG}
r^{\rm{V}}_2&=&r^{\rm{\overline{MS}}}_2-4a^{\rm{\overline{MS}}}_2-2a^{\rm{\overline{MS}}}_1r^{\rm{V}}_1 , \\ \label{r3VG}
r^{\rm{V}}_3&=&r^{\rm{\overline{MS}}}_3-4a^{\rm{\overline{MS}}}_3-3a^{\rm{\overline{MS}}}_1r^{\rm{V}}_2-(2a^{\rm{\overline{MS}}}_2+(a^{\rm{\overline{MS}}}_1)^2)r^{\rm{V}}_1 
\end{eqnarray}
and the  numerical values of these   
coefficients, namely    
\begin{eqnarray}
\label{r1V}
r^{\rm{V}}_1&=&2.59972n_f-9.5620 \; , \\ \label{r2V}
r^{\rm{V}}_2&=&0.50749n^2_f+113.6320n_f-2054.140-26.4435\delta_f \; , \\ 
\label{r3V}
r^{\rm{V}}_3&=&3.05815n^3_f-144.9455n^2_f+3455.279(2)n_f-20387.90(1.17) \\ 
\nonumber
&-&(39.0881n_f+701.466)\delta_f 
\end{eqnarray}
The errors in the values of the $n_f$ and $n_f^0$-terms in  
Eq. (\ref{r3V}) arise from the numerical errors in the  values of the 
$n_l$ and  $n_l^0$-dependent constituents   $a_3^{(1)}$ and $a_3^{(0)}$ 
of the coefficient 
$a_3^{\rm{\overline{MS}}}$ defined in   Eq.(\ref{a31}) and Ref.(\ref{a30}), which enter 
into the definition of  $r^{\rm{V}}_3$   through Eq.(\ref{r3VG}).

\subsection{The comparison of the fourth order  V- ,  $\rm{\overline{MS}}$- and mMOM-scheme\\ approximations for the $e^+e^-$  R-ratio.}  

As the start of the study of the scheme and energy dependence of the  
$e^+e^-$-annihilation R-ratio in different 
orders 
of PT in the case of applications of  three  different schemes  
we first present  in Table 2 the comparison of the following 
from Eqs. (\ref{r1V})-(\ref{r3V}) and Eqs. (\ref{r1})-(\ref{r2})   
numerical expressions for three  PT  coefficients  
in the V and  $\rm{\overline{MS}}$ scheme 
with the numerical expressions  of the  same  
coefficients, obtained  in the Landau-gauge version of the mMOM scheme 
in Ref. \cite{Gracey:2014pba}.

\begin{center}
\begin{tabular}{|c|c|c|c|c|c|c|c|c|c|}
\hline
\multicolumn{10}{|c|}{\textbf{ The  numerical coefficients of the R-ratio in different schemes}} \\
\hline 
$n_f$ & $r^{\rm{V}}_1$ & $r^{\rm{\overline{MS}}}_1$ & $r^{\rm{mMOM}}_1$ & $r^{\rm{V}}_2$ & $r^{\rm{\overline{MS}}}_2$ & $r^{\rm{mMOM}}_2$ & $r^{\rm{V}}_3$ & $r^{\rm{\overline{MS}}}_3$ & $r^{\rm{mMOM}}_3$ \\
\hline
1 & -6.9622 & 29.9265 & -21.9622 & -1966.444 & -528.346 & -1575.567 & -17815.06$\pm$1.17 & -36956.12 & -13190.55 \\
\hline
2 & -4.3625 & 28.0818 & -19.3625 & -1830.134 & -584.994 & -1467.688 & -14188.58$\pm$1.17 & -31536.11 & -8632.68 \\
\hline
3 & -1.7628 & 26.2371 & -16.7628 & -1708.676 & -658.171 & -1374.660 & -11244.00$\pm$1.17 & -27361.22 &  -4748.58 \\
\hline 
4 & 0.8368  & 24.3924 & -14.1631 & -1602.069 & -747.876 & -1296.483 & -9033.31$\pm$1.17 & -24310.08 & -1590.24 \\
\hline
5 & 3.4366 & 22.5477 & -11.5634 & -1475.696 & -819.494 & -1198.540 & -6434.41$\pm$1.17 & -20591.03 & 1575.00 \\
\hline
6 & 6.0363 & 20.7029 & -8.9637 & -1369.944 & -913.410 & -1121.218 & -4775.30$\pm$1.17 & -18149.16 & 3873.49 \\
\hline
\end{tabular}
\vspace{0.2cm}\\
{Table 2. The comparison of the numerical values of  the 
known coefficients for the $e^{+}e^{-}$-annihilation R-ratio 
in the V, $\rm{\overline{MS}}$ and in the Landau-gauge 
version of the  mMOM scheme.}
\end{center}
Note that
the values of the   coefficients 
 $r_i^{\rm{\overline{MS}}}$ ,  $r_i^{\rm{V}}$ and $r_i^{\rm{mMOM},\lambda=0}$
with i=2,3 
are {\it  negative}  for any number of $n_f$, apart from the 
case of  $r_3^{\rm{mMOM},\lambda=0}$ value  at $n_f=5,6$. 
 In the 
$\rm{\overline{MS}}$ scheme this  feature is  
related to the manifestation in the expressions for  
$r_2^{\rm{\overline{MS}}}$ and 
$r_3^{\rm{\overline{MS}}}$ 
of the effects proportional to $\pi^2$, which arise from  analytical continuation  to the Minkowskian region 
of energies of the PT contributions  in  the rhs  of 
Eq. (\ref{def}) (for a detailed explanation see  
e.g. Ref. (\cite{Kataev:1995vh}). The 
negative values of  the V-scheme 
coefficients are also related to these kinematic  $\pi^2$  effects, 
but the numerical  difference with  the negative values of 
$r_2^{\rm{\overline{MS}}}$ and $r_3^{\rm{\overline{MS}}}$-terms is related  
to  the numerical values of the  additions contributions to 
$r_2^{\rm{V}}$ and $r_3^{\rm{V}}$-terms. Note  that  in the case of 
$r_2^{\rm{V}}$ they  are 
negative (see Eq. (\ref{r2VG}))
but in the case of $r_3^{\rm{V}}$ they  are positive due to interplay among the 
third huge positive contribution  to  Eq. (\ref{r3VG}) and  other  
negative contributions to the same equation. Note also that 
the values of $r_2^{\rm{V}}$ are  very closed to $r_2^{\rm{mMOM}}$, but this 
feature  does not remain  at the fourth order of PT. 
   
We now plot the energy and scheme dependence of the  
next-to-leading order (NLO), next-to-next-to-leading order (NNLO) and 
next-to-next-to-next-to-leading order (N$^3$LO) approximations   for  the 
function   $ r(s)=R(s)/(3\sum_f Q^2_f)-1$.  It   
depends on   $s=q^2$, where $s$ is 
measured in ${\rm GeV}^2$. The first three  plots are presented in Fig.  1 
for the energy region above the threshold of charmonium production and 
below the  threshold of the bottomonium production, i. e. in the 
region where   $n_f=4$ numbers of active flavours are contributing 
to the expression for $r(s)$. In Fig. 2 the 
scheme dependence of the NLO, NNLO and N$^3$LO approximations of the same 
function are presented in region with   $n_f=5$ numbers 
of active flavoures.  More definitely, we consider 
the energy region above the  threshold of bottomonium 
production and up to the energies  $s=900~{\rm GeV}^2$,
where the subprocess
$e^+e^-\rightarrow Z^0\rightarrow {\rm{hadrons}}$, which starts to  dominate near  
the beginning of the  left 
shoulder of the direct manifestation of $Z^0$-boson in the  $e^+e^-$-
collisions, can be safely neglected. 

The energy dependence of  coupling constant  
$a_s=\alpha_s^{\rm{\overline{MS}}}/(4\pi)$ of the NLO, NNLO  
approximations  
of the PT expansions of  the $e^+e^-$-annihilation  
ratio $R(s)$  in the  $\overline{\rm{MS}}$ scheme, 
 which is presented in Eq. (\ref{MSbar}), 
is defined through the powers of logarithmic terms 
$\rm{L}=\ln(s/\Lambda^{{(n_f)} 2}_{\rm{\overline{MS}}})$-   
as 
\begin{equation}
\label{NLO}
a_s^{\rm{NLO}}=\frac{1}{\beta_0
\rm{L}} -\frac{\beta_1 \ln(\rm{L})}{\beta_0^3 \rm{L}^2}
\end{equation}
\begin{equation}
a_s^{\rm{NNLO}}=a_s^{\rm{NLO}}+\Delta a_s^{\rm{NNLO}} 
\end{equation} 
where 
\begin{equation}
\label{NNLO}
\Delta a_s^{\rm{NNLO}}=\frac{1}{\beta_0^5 \rm{L}^3}[\beta_1^2 \ln^2 (\rm{L})
-\beta_1^2 \ln(\rm{L}) +\beta_2\beta_0-\beta_1^2]
\end{equation}
At the fourth N$^3$LO, first studied in Ref. \cite{Chetyrkin:1997sg}, 
one has  
\begin{equation}
a_s^{\rm{N^3LO}}=a_s^{\rm{NNLO}}+\Delta a_s^{\rm{N^3LO}}
\end{equation}
where the additional correction reads: 
\begin{equation}
\label{N3LO}
\Delta a_s^{\rm{N^3LO}}=\frac{1}{\beta_0^7 \rm{L}^4}\bigg[\beta_1^3 \bigg(-\ln^3 (\rm{L})
+\frac{5}{2}\ln^2 (\rm{L})
+2\ln({L})-\frac{1}{2}\bigg)
-3\beta_0\beta_1\beta_2 \ln({L})
+\beta_0^2\frac{\beta_3}{2}\bigg]~.
\end{equation}

\begin{figure}[t]
\includegraphics[width=0.55\textwidth]{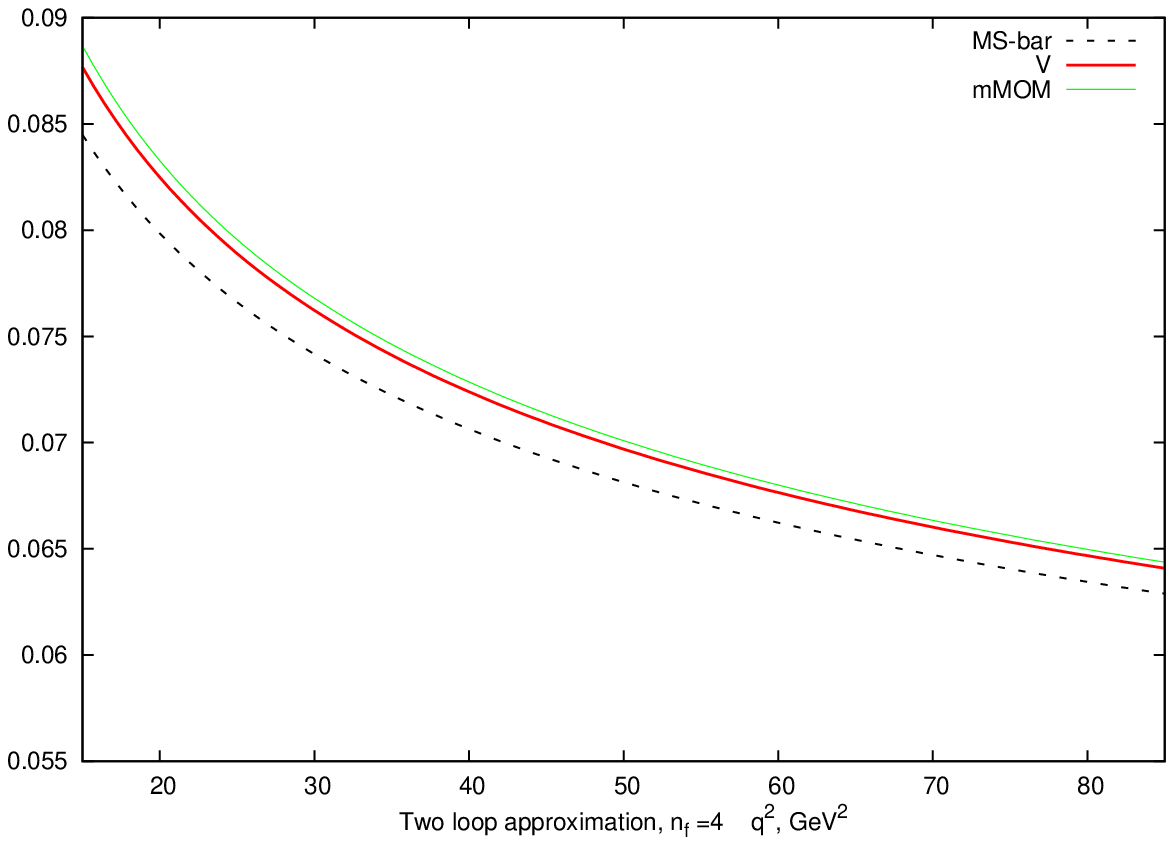}~~
\includegraphics[width=0.55\textwidth]{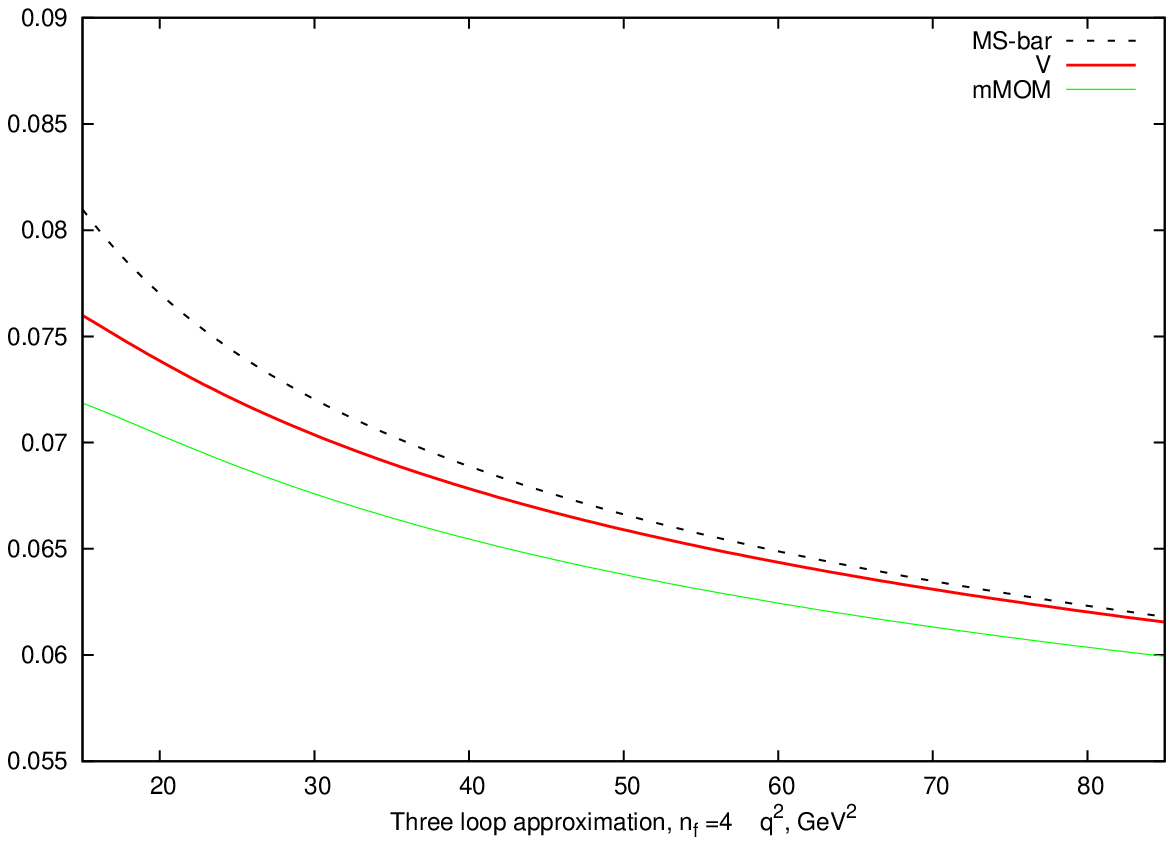}
\includegraphics[width=0.55\textwidth]{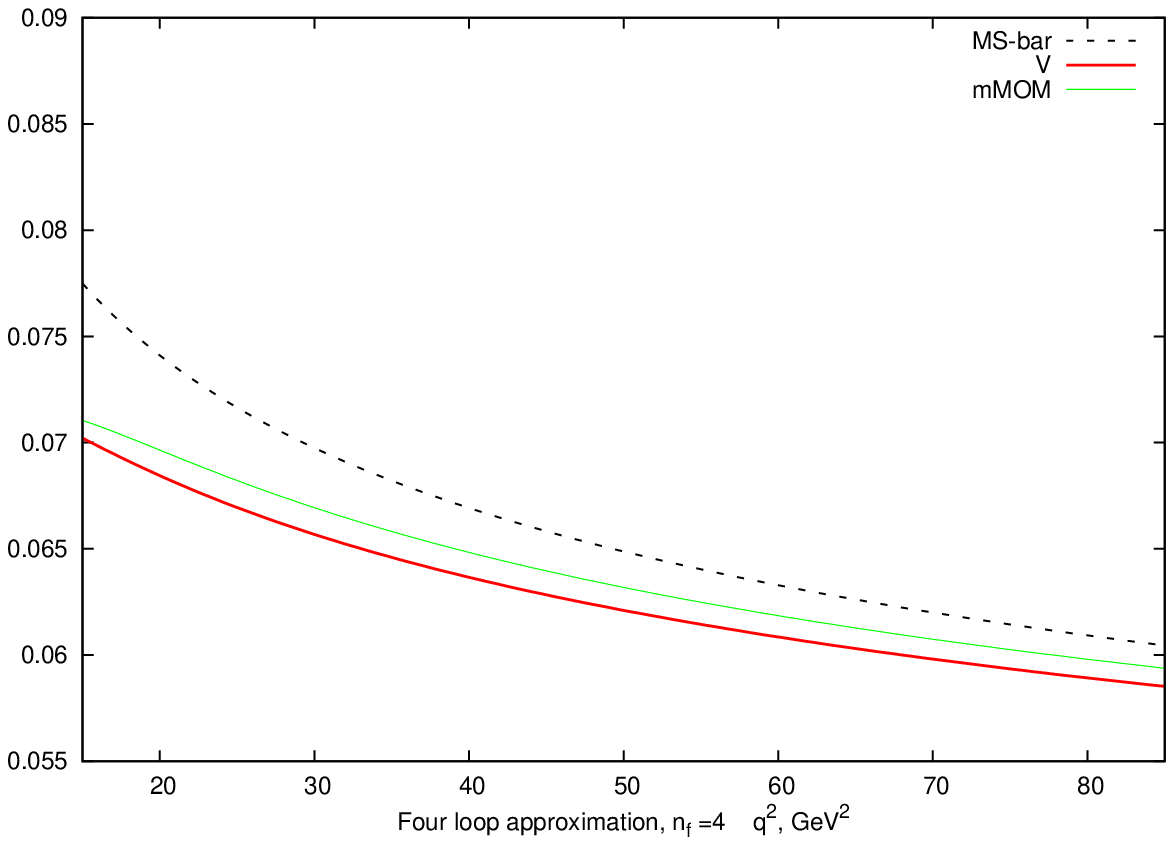}~~
\caption{\footnotesize Scheme dependence of  the NLO (left), NNLO (right) and N$^3$LO (bottom) approximations 
for the e$^{+}e^{-}$ characteristic  $r(q^2)=R(q^2)/(3\sum_f Q^2_f)-1$ 
in the case of $n_f$=4 numbers of active flavours. 
The dashed black curve depicts the variations of the 
${\rm{\overline{MS}}}$ approximants. 
The solid (green) line demonstrate the variations of  
the  ${\rm{mMOM}}$-scheme results, while 
the solid red line shows  the V-scheme results.} 
\end{figure}
In the numerical form the  expressions for the 
$\rm{\overline{MS}}$ $\beta$-functions coefficients  
$\beta_i$  in Eqs. (\ref{NLO})-(\ref{N3LO}) are defined in Eqs. (\ref{b0-b1}), 
(\ref{B1}) and Eqs.(\ref{b2n}),(\ref{b3n}) respectively.  
For the concrete numbers of 
$n_f$ flavours  their values  are given in Table 1. Note that in the analysis 
of Ref.\cite{Gracey:2014pba} the same  expansion was used
for the numbers of active flavours $n_f$=$5$ and $n_f$=$6$ and 
for the value of $\Lambda_{\rm{\overline{MS}}}=500$ MeV, which did not vary from 
order to order of the $\rm{\overline{MS}}$-scheme perturbative  expressions considered in Ref. \cite{Gracey:2014pba}. In the process 
of obtaining our results, presented in Figs. 1 and 2, and  keeping in mind 
physical motivations, discussed above,  
we used $n_f$=$4$ and  $n_f$=$5$.   

\begin{figure}[t]
\includegraphics[width=0.55\textwidth]{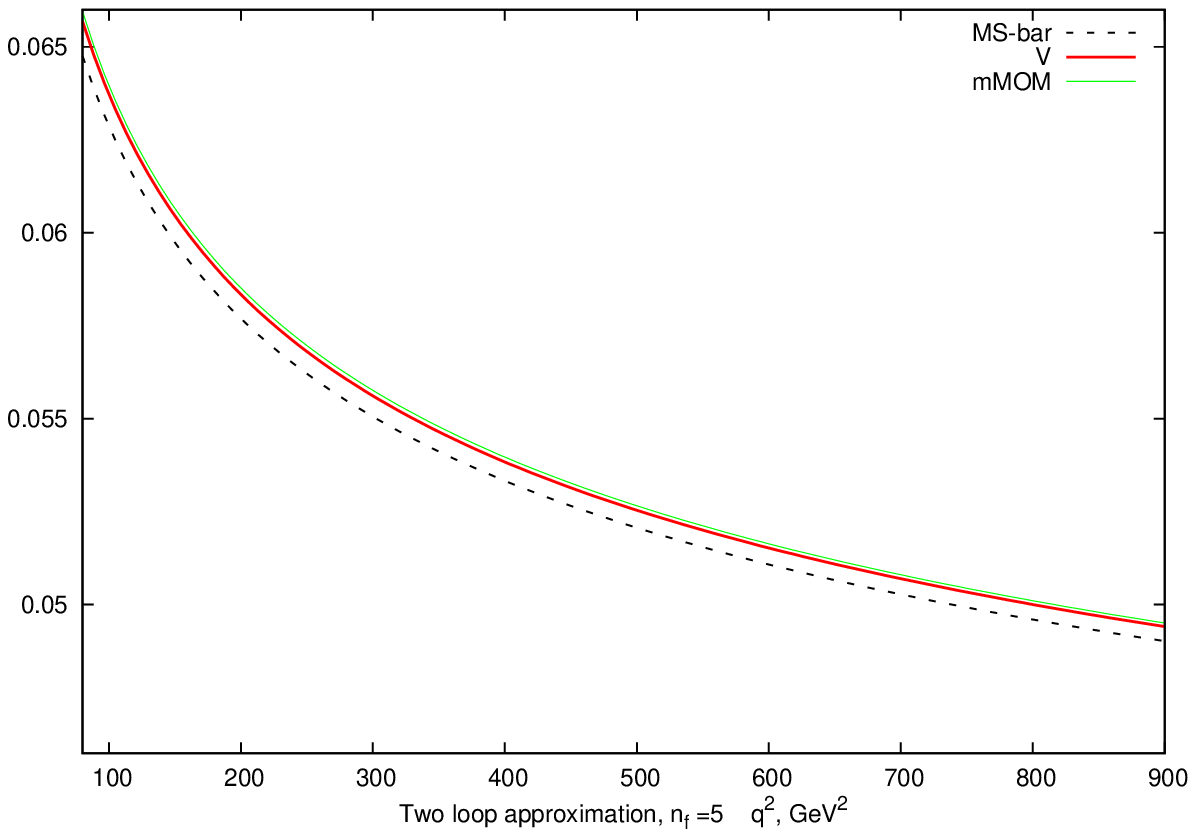}~~
\includegraphics[width=0.55\textwidth]{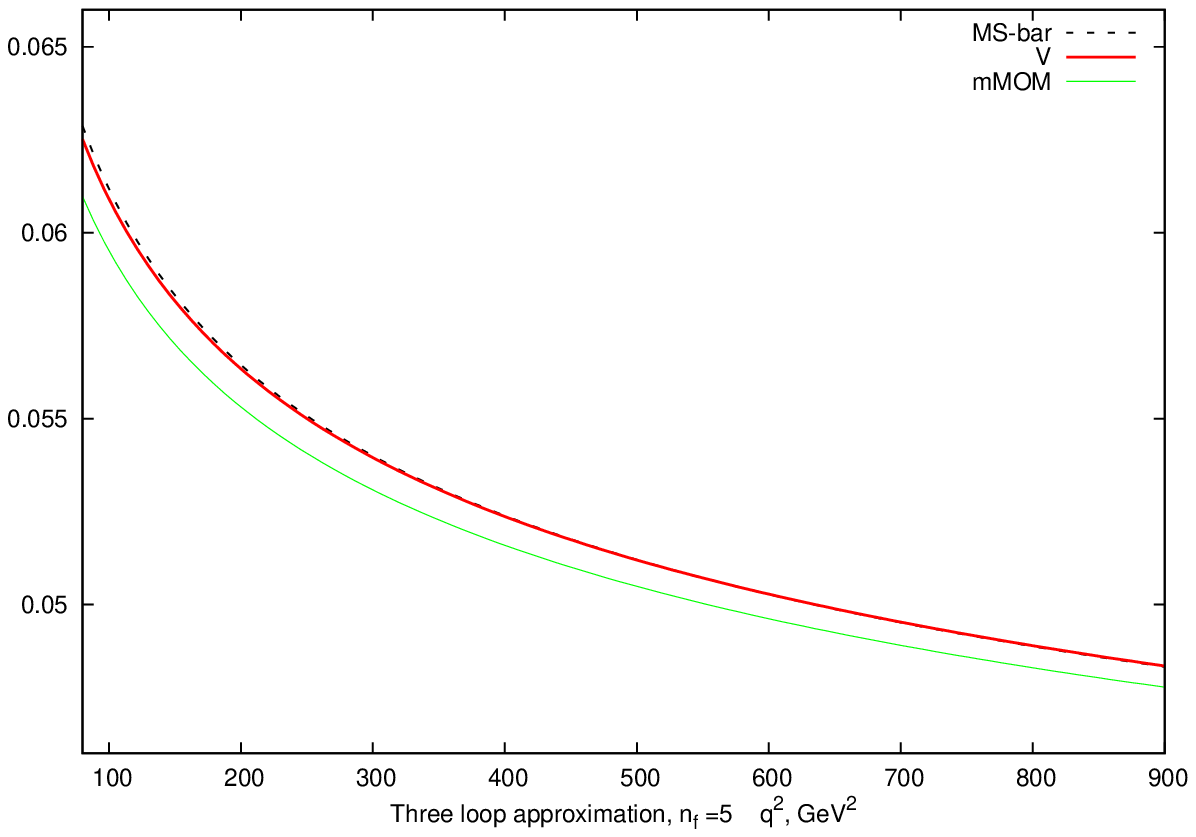}
\includegraphics[width=0.55\textwidth]{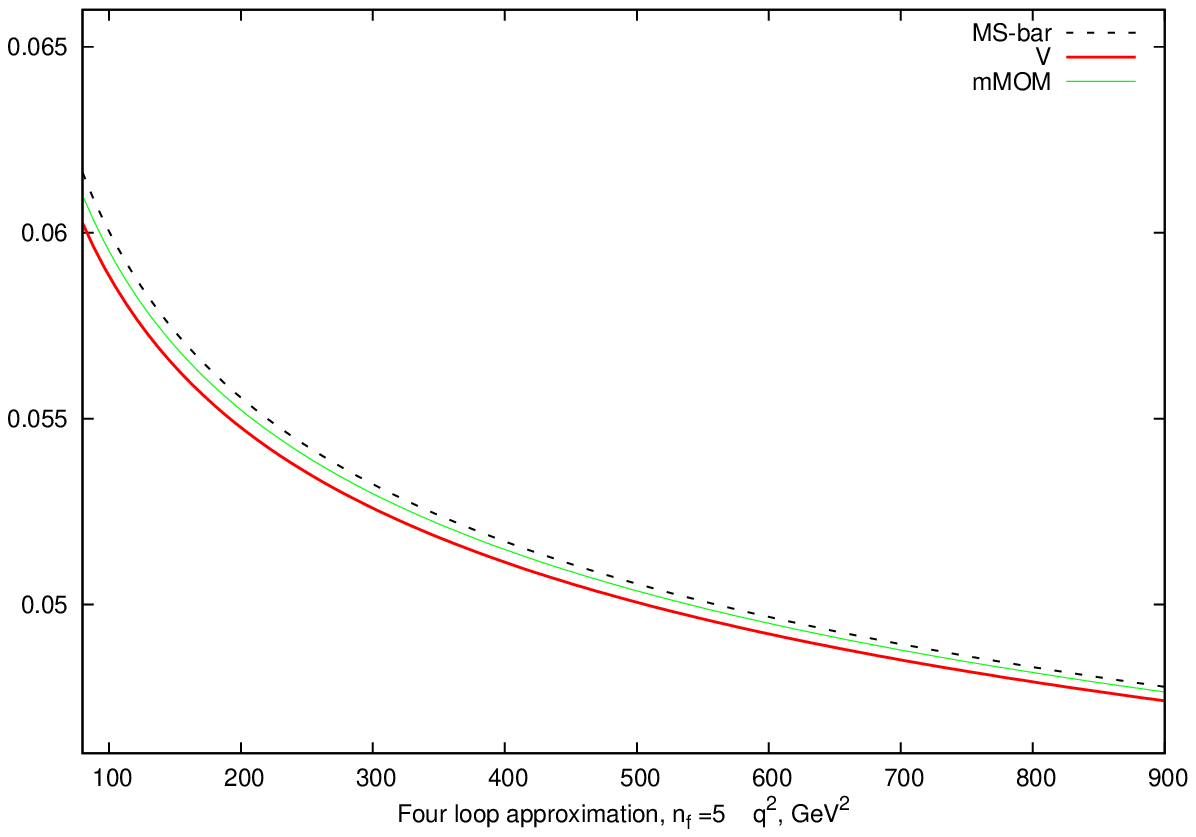}~~
\caption{ \footnotesize  Scheme dependence of  the NLO (left), NNLO (right) and N$^3$LO (bottom) approximations to 
$r(q^2)$ are presented for $n_f$=$5$ numbers of active flavours. 
The variation of the ${\rm{\overline{MS}}}$-, mMOM- and V-scheme results is 
indicated  by the three curves as in Fig.1 }
\end{figure}

Contrary to the studies of Ref. \cite{Gracey:2014pba}
the  values of the 
parameters $\Lambda^{(n_f)}_{\rm{\overline{MS}}}$ ,   $\Lambda^{(n_f)}_{\rm{mMOM}}$ 
and parameter $\Lambda^{(n_f)}_{\rm{V}}$ (that is new to this work)
were not fixed,
but  depend on the choice of  both $n_f$ and the  order 
of approximations. The concrete  results for the values 
of the parameters used are presented in Table 3.

\begin{center}  
 \begin{tabular}{|c|c|c|c|c|}
\hline
\multicolumn{5}{|c|}{\textbf{ The  numerical values of the $\Lambda_{\rm{QCD}}$ in different schemes, \rm{MeV}}} \\
\hline 
$n_f$ & $  \text{the order of approximation $\nu$ }$ & $\Lambda^{(n_f)}_{\overline{\rm{MS}}}$ & $\Lambda^{(n_f)}_{\rm{V}}$ & $\Lambda^{(n_f)}_{\rm{mMOM}}$ \\
\hline
 4 & 2 & $\;\;\;\;\;\;\;\;\;\;$ 350 $\;\;\;\;\;\;\;\;\;\;$ & $\;\;\;\;\;\;\;\;\;\;$ 500 $\;\;\;\;\;\;\;\;\;\;$ & $\;\;\;\;\;\;\;\;$ 625 $\;\;\;\;\;\;\;\;$ \\
\hline
4 & 3 & 335 & 475 & 600 \\
\hline
4 & 4 & 330 & 470 & 590 \\
\hline
5 & 2 & 250 & 340 & 435  \\
\hline 
5 & 3 & 245 & 335 & 430 \\
\hline
5 & 4 & 240 & 330 & 420 \\
\hline
\end{tabular}
\vspace{0.2cm}\\
{Table 3. The dependence of the parameters,  used for  
getting the results of Figs. 1 and 2 from the $n_f$, 
$\nu$ (order of approximation), and from the choice of the scheme.}
%The relations of $\Lambda^{(n_f)}_{\rm{V}}$ and 
%$\Lambda^{(n_f),\lambda=0}_{\rm{mMOM}}$ with $\Lambda^{(n_f)}_{\rm{\overline{MS%}}}$
%are given below.} 
\end{center}

In the cases of $n_f$=$4$ numbers of active flavours and $\nu=2,3,4$  
the values  for $\Lambda^{(n_f=4)}_{\rm{\overline{MS}}}$  given in   
Table 3  are fixed from the 
results of the fits
fits of the Fermilab Tevatron  experimental data for the 
$xF_3$ structure function of the  neutrino-nucleon deep-inelastic scattering 
process at the $\rm{N^{(\nu-1)}LO}$ of the  theoretical PT results,  performed in Ref.\cite{Kataev:2001kk}.   In the case of $n_f=5$   
the values of $\Lambda^{(n_f=5)}_{\rm{\overline{MS}}}$ at $\nu=2,3,4$ were 
obtained in Ref. \cite{Kataev:2009ns} from the related results for   $\Lambda^{(n_f=4)}_{\rm{\overline{MS}}}$
using the  the NLO, NNLO and $\rm{N^{3}LO}$ matching 
conditions, evaluated at the NNLO in Ref. \cite{Bernreuther:1981sg}
and \cite{Larin:1994va} and at the $\rm{N^{3}LO}$ in 
Ref. \cite{Chetyrkin:1997sg}. The  matching point in these  conditions was fixed by the    
on-shell b-quark mass values, extracted     
 at different orders of PT from the analysis of 
heavy quarkonium spectrum while taking into account the Pade estimated  
value of  the coefficient $a_3$ from  Eq. (\ref{a3}), obtained  in Ref. \cite{Chishtie:2001mf}. These Pade  estimates  
turned out to be in 
satisfactory  agreement with the results of direct calculations of the value of  $a_3$ obtained  
later   
(see Refs. \cite{Smirnov:2008pn}, \cite{Smirnov:2009fh}, \cite{Anzai:2009tm}).
In view of reliability of the results of  Ref. \cite{Penin:2002zv}
we may safely use the
values for    
$\Lambda^{(n_f=5)}_{\rm{\overline{MS}}}$ from Table 3  for transforming them to 
the values of the scale parameters 
$\Lambda^{(n_f=5)}_{\rm{mMOM}}$ and $\Lambda^{(n_f=5)}_{\rm{V}}$ {\it in particular}.

In general the scale 
parameters  $\Lambda^{(n_f)}$ of  the $\rm{\overline{MS}}$ , V and $\rm{mMOM}$ 
schemes  considered  in Table 3  are related by  the following equations: 
\begin{equation}
\label{Lrelations}
\Lambda^{{(n_f)}2}_{\rm{V}}=\Lambda^{{(n_f)}2}_{\overline{\rm{MS}}}\rm{exp}
[a^{\overline{\rm{MS}}}_1(n_f)/\beta_0(n_f)]~,~ 
\Lambda^{{(n_f)}2}_{\rm{mMOM}}=\Lambda^{{(n_f)}2}_{\overline{\rm{MS}}}\rm{exp}
[(r^{\overline{\rm{MS}}}_1(n_f)-r^{\rm{mMOM}}_1(n_f))/4\beta_0(n_f)]~.
\end{equation}
They are  derived by means of  the ECH approach. We 
used these  expressions  to get 
in Table 3 
the numerical values 
of $\Lambda^{(n_f)}_{\rm{V}}$ and  $\Lambda^{(n_f)}_{\rm{mMOM}}$ from the
  results described above for 
$\Lambda^{(n_f)}_{\overline{\rm{MS}}}$. Combining them with the numerical 
values for the  coefficients $\beta_2^{\rm{V}}(n_f)$ , $\beta_3^{\rm{V}}(n_f)$ 
and  $\beta^{\rm{mMOM},\lambda=0}_2(n_f)$ ,  $\beta^{\rm{{mMOM},\lambda=0}}_3(n_f)$ 
in the  analogs of Eqs. (\ref{NLO}), (\ref{NNLO}) and Eq.(\ref{N3LO}),
and taking into account the  expressions for the  
coefficients $r_i$ in 
$r(q^2)=R(q^2)/(3\sum_f Q^2_f)-1$ in  three different schemes, we   
plot  in Figs.1 and 2 the energy dependence of $r(q^2)$ in three different orders of PT 
and three different schemes, namely $\rm{\overline{MS}}$, V and mMOM schemes
in the  case of $n_f=4$ and $n_f=5$ respectively. 

\newpage
 
\subsection{Discussions of the results.}

Considering now the plots of Figs.1 and  2 we may conclude that in all 
cases the PT approximants for the function $r(s)$  related to the $e^+e^-$-annihilation 
R-ratio 
are converging in all schemes. In the $\rm{\overline{MS}}$ scheme the  
rate of convergence of the related PT approximants is better than 
in the V scheme and mMOM scheme. At the NLO the results of the V scheme are closer to the mMOM ones than to the results 
obtained in  $\rm{\overline{MS}}$ scheme, while at the NNLO the situation is reversed -- the V-scheme approximations are closer to 
the  $\rm{\overline{MS}}$-ones, while the application of the mMOM scheme puts a lower bound on the theoretical expression 
for $r(s)$. However, at the N$^{3}$LO the lower theoretical bound on the energy dependence  of  $r(s)$ is changed again and the lower bound 
is now obtained within V scheme. The comparison of three approximants for $r(s)$ in the case 
of consideration of the V-sche me results supports the conclusion, made in Sec. IV. C, that the PT approximants in  the V scheme have less regular 
behaviour than the $\rm{\overline{MS}}$ ones. 
The results of Table 2 demonstrate the positive feature of taking into account 
$O(\alpha_s^4)$-corrections to $e^+e^-$ annihilation R-ratio in all three 
schemes. Indeed, the  scheme dependence of the 
expression for the $e^+e^-$ ratio is drastically decreased at this level. 
This is the positive message, which supports the  work presented above 
on the inclusion of the  $O(\alpha_s^4)$ correction in the theoretical 
approximations in the $\rm{\overline{MS}}$, mMOM and V schemes.

\section{The four-loop QED result for the RG $\beta$ function 
in the    $\rm{V}$ scheme}

Consider now the case of QED with N types  of identically charged  leptons.   We will use  the results 
of Sec. \ref{betaV:QCD}  for the    
the fourth-order PT approximation of the  RG  $\rm{V}$-scheme $\beta$-function of  the 
$SU(N_c)$ colour  gauge group theory. 
Fixing   the $SU(N_c)$ group weights in Eqs. (\ref{beta0}), (\ref{beta1}), (\ref{b2V}) and   (\ref{b3V})  
as  $C_A=0$, $C_F=1$, $T_F=1$, $d^{abcd}_A=0$, $d^{abcd}_{F}=1$, $N_A=1$ and $n_f=N$,
we obtain the following   four-loop semi-analytical 
expression  for the RG $\beta$ function in QED in the ${\rm V}$ scheme: 
\begin{eqnarray}
\label{new2}
&&\beta^{\rm{V}}_{QED}(a_V)=\frac{4}{3}Na_{\rm{V}}^2+4Na_{\rm{V}}^3
+\bigg(-2N + (\frac{64}{3}\zeta(3)-\frac{184}{9})N^2\bigg)a_{\rm{V}}^4+ \\
\nonumber 
&+&\bigg(-46 N+(104 +\frac{512}{3}\zeta(3)-\frac{1280}{3}\zeta(5)-
\frac{8}{3}\cdot56.83(1))N^2
+(128-\frac{256}{3}\zeta(3))N^3\bigg)a_{\rm{V}}^5 
+O(a_{\rm{V}}^6) 
\end{eqnarray}
where $a_{\rm{V}}=\alpha_{\rm{V}}/4\pi$ and $N$ is the  number of leptons. 
Comparing this result with  the four-loop 
approximation of  the QED 
$\beta$ function in the MOM scheme, i. e. of  the Gell-Man--Low  
$\Psi$ function, namely with  
\begin{eqnarray}
\label{QEDPsi} 
&&\Psi(a_{\rm{MOM}})=\frac{4}{3}Na_{\rm{MOM}}^2+4Na_{\rm{MOM}}^3+
\bigg(-2N + (\frac{64}{3}\zeta(3)-\frac{184}{9})N^2\bigg)a_{\rm{MOM}}^4
\\ \nonumber 
&+&\bigg(-46 N+(104 +\frac{512}{3}\zeta(3)-\frac{1280}{3}\zeta(5))N^2+ 
(128-\frac{256}{3}\zeta(3))N^3\bigg)a_{\rm{MOM}}^5+O(a_{\rm{MOM}}^6) 
\end{eqnarray}
where $a_{\rm{MOM}}=\alpha_{\rm{MOM}}/4\pi$, we conclude that in spite of 
identical agreement at the third order of PT \footnote{This observation was made and 
used in the unpublished 
work of   A.L.Kataev and A.V. Garkusha, see Ref. \cite{AVG} as well.},
the general  expressions for the RG  QED $\beta$ function in these 
two different schemes are not the same. They start to differ from the fourth order of PT 
due to contributing to the $O(a_{\rm{V}}^5)$ coefficient of the 
$\beta^{\rm{V}}$-function of the additional light-by-light-type scattering diagrams, which appear in the QED analog of the coefficient $a_3^{(1)}$ in the 
$\rm{\overline{MS}}$ scheme, given in Eq. (\ref{a31}).
They enter in the definition of the $N^2$-term 
of the $\beta_3^{\rm{V}}$ coefficient of the V-scheme QED $\beta$-function through Eq. (\ref{t3v}).

It is possible to clarify what kind of N-dependent high-order  
coefficients of the following expression of the 
QED $\beta$ function in the V scheme 
\begin{equation} 
\label{betaVN}
\beta^{\rm{V}}(a_{\rm{V}})=\sum\limits_{i= 0}^{\infty}\beta_i^{\rm{V}}
\bigg(\frac{\alpha_{{\rm{V}}}}{4\pi}\bigg)^{i+2} =
\beta^{\rm{V}[1]}_0N\bigg(\frac{\alpha_{{\rm{V}}}}{4\pi}\bigg)^2+\sum\limits_{i=1}^{\infty}\sum\limits_{l=1}^i\beta^{\rm{V}[l]}_iN^l\bigg(\frac{\alpha_{{\rm{V}}}}{4\pi}\bigg)^{i+2} 
\end{equation}
will  also receive additional contributions and what kind of the N-dependent 
coefficients  of  the QED 
$\beta^{\rm{V}}$ function will coincide with the similar expressions   
for the $\Psi$ function, which we will define as  
\begin{equation}
\label{PsiVN}
\Psi(a_{\rm{MOM}})=\Psi^{[1]}_0N\bigg(\frac{\alpha_{\rm{MOM}}}{4\pi}\bigg)^2+
\sum\limits_{i=1}^{\infty}\sum\limits_{l=1}^i\Psi^{[l]}_iN^l\bigg(\frac{\alpha_{\rm{MOM}}}{4\pi}\bigg)^{i+2}  .
\end{equation}
Using the analogs of Eq. (\ref{t2}) and  (\ref{t3v}), which 
can be derived using the considerations of Ref. \cite{Kataev:1995vh},
we arrive at the following relations:  
\begin{equation}
\label{exrtaN}
\beta_i^{{\rm{V}[l]}}=\Psi_{i}^{[l]}+\Delta\beta_i^{{\rm{V}[l]}}
\end{equation} 
where extra terms  $\Delta\beta_i^{{\rm{V}[l]}}$ in the N-dependent 
contributions to the coefficients of the QED $\beta^{V}$ function   
appear in the following region of indexes $[i,l]=[i\geq 3, 2\leq l\leq i-1]$.

In the cases of  $[i,l]=[i\geq 3,l=1 ~\text{or}~  i]$  
the proportional to $N^{[l]}$ coefficients of the $\beta^{\rm{V}}$ and 
$\Psi$ function, defined in Eq. (\ref{betaVN}) and (\ref{PsiVN}), are the same. In the case of i=3, which corresponds to the totally known for the moment
fourth order results, these identical coefficients are proportional
to N and $N^3$. At the third order the proportional to N-term
was analytically evaluated in Ref. \citep{Rosner:1967zz}. At the fourth
order of PT the proportional to N and $N^3$ terms were
evaluated in Ref. \citep{Gorishnii:1990kd}. For  $i=4$ the terms under discussion  can be obtained from the 
results of Ref.\cite{Baikov:2012zm} and read  
\begin{eqnarray}
\label{N}
&&\beta_4^{{\rm{V}[1]}}=\Psi_4^{[1]}=
\frac{4157}{6}+128\zeta(3) \\ 
\label{N4}
&&\beta_4^{{\rm{V}[4]}}=\Psi_4^{[4]}= 
-\frac{8756}{9}+\frac{3584}{9}\zeta(3)+\frac{5120}{9}\zeta(5)
\end{eqnarray}
Note that this result from Ref. \cite{Baikov:2012zm} is in agreement 
with the multiloop expression for this particular contribution to the 
Gell-Man--Low function, evaluated in Ref. \cite{Broadhurst:1992si}
up to 20 loops analytically and numerically up to 100 loops.
The scheme-independence of the linear-in-N-contribution 
to Eqs. (\ref{betaVN}) and  (\ref{PsiVN}) is the consequence of the 
{\it conformal symmetry} property, which is valid in QED in the perturbative 
quenched approximation (for the recent detailed study see 
Ref. \cite{Kataev:2013vua}).

In the numerical form the  scheme-dependent coefficients of the  
$\beta^V_{\rm{QED}}$-function read:
\begin{eqnarray}
\beta_2^{\rm{V}}&=&-2N+5.19943N^2 \\
\beta_3^{\rm{V}}&=&-46N+284.818(26)N^2-25.42447N^3
\end{eqnarray}
The  analogous expressions for the three- and 
four-loop coefficients of the QED   $\beta$ function in the  
$\overline{\rm{MS}}$ scheme follow  from the 
analytical results of Ref. \cite{Gorishnii:1990kd} and have the following 
form  
\begin{eqnarray}
\label{eqnarry}
\beta_2^{\overline{\rm{MS}}}&=&-2N+4.88888N^2 \\
\beta_3^{\overline{\rm{MS}}}&=&-46N +82.9753N^2+5.06995N^3
\end{eqnarray}
The numerical expressions for the analogous  
coefficients of the  $\Psi$ function
(or the QED $\beta$ function in MOM scheme), which we obtain 
from the same work of Ref. \cite{Gorishnii:1990kd}, are  
\begin{eqnarray}
\Psi_2&=&-2N+5.19943N^2 \\
\Psi_3&=&-46N+133.2714N^2-25.42447N^3
\end{eqnarray}
Note once more that the first three     
coefficients  
of the  $\beta^{\rm{V}}$ function and of the $\Psi$ function 
are the same and start to differ from the fourth order of PT
in the following way  
\begin{equation}
\label{extra}
\beta_{3}^{\rm{V}}=\Psi_3-151.54(2)N^2
\end{equation}
This additional contribution arises from the light-by-light-type  
scattering contribution, which is typical to the V scheme. 

For completeness we present the QED expressions for the $O(\alpha^5)$
approximations for the $\Psi$ and $\beta^V_{QED}$ functions in the case 
of N=1 :  
\begin{eqnarray}
\Psi(a_{\rm{MOM}})&=&1.3333a_{\rm{MOM}}^2+4a_{\rm{MOM}}^3+3.1994a_{\rm{MOM}}^4-
153.8469a_{\rm{MOM}}^5+O(a_{\rm{MOM}}^6) \\ 
\beta^{V}_{QED}(a_{\rm{v}})&=&1.3333a_{\rm{V}}^2+4a_{\rm{V}}^3+3.1994a_{\rm{V}}^4-
305.3936(266)a_{\rm{V}}^5 +O(a_{\rm{V}}^6)
\end{eqnarray}

One can observe that even for N=1  
the numerical  effect of light-light-scattering 
contribution, which is typical for the V-scheme (see Eq. (\ref{extra}) ),
is rather sizable and almost equals to  the whole value of the
other term in the expression of Eq. (\ref{extra}).

\section{Conclusion}
In this work we consider the definition of  
the gauge-independent RG QCD $\beta$ function in the $\rm{V}$ scheme. 
 Using higher-order corrections to the static potential of the 
quark-antiquark interaction and $\beta$ function in 
$\overline{\rm{MS}}$ scheme, we  compute the fourth term of the  
PT expression  for  
the    $\beta$ function in $\rm{V}$ scheme  in the 
general case of  $\rm{SU(N_c)}$ group in the semi-analytical 
term.  Our guess of possible expressions of the corresponding numerical 
contributions through concrete transcendental numbers is made.   
 The 
comparison of the numerical expressions of the scheme-dependent coefficients of the $\beta^{\rm{V}}$ function of QCD 
with the similar coefficients of the QCD $\beta$ function in the $\rm{\overline{MS}}$ and $\rm{mMOM}$ scheme in the Landau 
gauge are presented. The indication that the 
structure of the PT series for the  effective $\beta$ function in the 
$\rm{V}$ scheme has non-regular asymptotic  behaviour and differs from 
the asymptotic PT for the $\beta$ function in the 
$\rm{\overline{MS}}$ scheme are presented.  The results obtained 
in the V scheme 
are used to study the scheme dependence of the 
$O(\alpha_s^4)$ approximation for the $e^+e^-$ annihilation R-ratio
in the energy region above the thresholds of production of the charmonium 
states. 
The conclusion is made that the comparison between the fourth-order 
expressions  for the $e^+e^-$ annihilation R-ratio, obtained in the 
$\rm{\overline{MS}}$ schemes, in the Landau-gauge variant  of the   
mMOM scheme and in the   gauge-independent V scheme leads to a drastic
decrease of the scheme dependence of the fourth-order perturbative QCD 
predictions for the case of $n_f=5$ numbers of active flavours 
in particular. Considering the QED limit of the  $\rm{SU(N_c)}$-group  
$\beta^{\rm{V}}$ function   
we observe that its perturbative expression is starting to differ from 
the perturbative expression for 
the Gell-Mann--Low $\Psi$ function from the level of 
the  $O(\alpha_{\rm{V}}^6)$-corrections.
The relations between coefficients of the QED  $\beta^{\rm{V}}$ function and the 
$\Psi$ function are presented in all orders of PT in the case 
of the N-types of identical leptons.  The conclusion that starting from the 
fourth-order perturbative approximation 
 two N-dependent terms in the  coefficients of 
the perturbative expansions  of the $\beta^{\rm{V}}$ and $\Psi$ functions
will always coincide is made. Theoretical reasons of this foundations are 
presented. 

\acknowledgments

The work  on phenomenologically oriented applications of  
the ${\rm{V}}$ scheme to the analysis of the fourth-order approximation 
of the total cross-section of the  $e^+e^-$ annihilation to hadrons process  
was   supported   by the  Russian Science Foundation Grant  
 N 14-22-00161. 
We wish to thank  S.J. Brodsky, D.G. Levkov and Y. Sumino 
for   useful questions and comments. 
%%%%%%%%%%%%%%%%%%%%%%%%%%%%%%%%%%%%%%%%%%%%%%%%%%%%%%%%%%%%%%%%%%%%%%%%%%%%%%%

%\begin{thebibliography}{[**]}

\newcommand{\noopsort}[1]{} \newcommand{\printfirst}[2]{#1}
\newcommand{\singleletter}[1]{#1} \newcommand{\switchargs}[2]{#2#1}

\end{document}